\documentclass[fleqn,usenatbib]{mnras}

\usepackage{newtxtext,newtxmath}
\usepackage[T1]{fontenc}
\usepackage{subfig}

\DeclareRobustCommand{\VAN}[3]{#2}
\let\VANthebibliography\thebibliography
\def\thebibliography{\DeclareRobustCommand{\VAN}[3]{##3}\VANthebibliography}
\usepackage{color,soul}
\definecolor{lightblue}{rgb}{.70,.95,1}
\sethlcolor{lightblue}

\usepackage{graphicx}	% Including figure files
\usepackage{chemformula}
\usepackage{float}
\usepackage{subfig}
\usepackage{bm}
\usepackage{hyperref}
\usepackage{gensymb}
\usepackage[most]{tcolorbox}
\usepackage{xcolor}
\usepackage{adjustbox}
\usepackage{multicol}
\usepackage{wrapfig}
\newcommand{\feh}{\ensuremath{\ch{[Fe/H]}} }
\newcommand{\cfe}{\ensuremath{\ch{[C/Fe]}} }

\title[On the dearth of CEMP stars in the Galactic bulge]{On the dearth of C-enhanced metal-poor stars in the Galactic bulge}

\author[G. Pagnini, S. Salvadori, M. Rossi et al.]{
G. Pagnini,$^{1}$\thanks{E-mail: giulia.pagnini@obspm.fr}
S. Salvadori,$^{2,3}$
M. Rossi,$^{2,3}$
D. Aguado,$^{2,3,4}$
I. Koutsouridou$^{2,3}$
and \'{A}. Sk\'{u}lad\'{o}ttir$^{2,3}$
\\
$^{1}$GEPI, Observatoire de Paris, PSL Research University, CNRS, Place Jules Janssen, 92195 Meudon, France\\
$^{2}$Dipartimento di Fisica e Astronomia, Universit\'{a} degli Studi di Firenze, via G. Sansone 1, 50019, Sesto Fiorentino, Italy\\
$^{3}$INAF/Osservatorio Astrofisico di Arcetri, Largo E. Fermi 5, I-50125, Firenze, Italy\\
$^{4}$Instituto de Astrof\'{\i}sica de Canarias,
              V\'{\i}a L\'actea, 38205 La Laguna, Tenerife, Spain
}
\date{Accepted XXX. Received YYY; in original form ZZZ}

\pubyear{2015}

\begin{document}
\label{firstpage}
\pagerange{\pageref{firstpage}--\pageref{lastpage}}
\maketitle

\begin{abstract}
The chemical fingerprints of the first stars are retained within the photospheres of ancient unevolved metal-poor stars. A significant fraction of these stellar fossils is represented by stars known as Carbon-Enhanced Metal-Poor (CEMP), $\cfe>+0.7$ and $\feh<-2$, which are likely imprinted by low-energy primordial supernovae. These CEMP stars are largely observed in the Galactic halo and ultra-faint dwarf galaxies, with values reaching $\rm [C/Fe]=+4.5$. The Galactic bulge is predicted to host the oldest stars, but it shows a striking dearth of CEMP stars with $\rm [C/Fe]\gtrsim +2.0$. Here we explore the possible reasons for this anomaly by performing a statistical analysis of the observations of metal-poor stars in combination with the predictions of $\Lambda$CDM models. We suggest that the dearth of CEMP stars with high \cfe is not due to the low statistics of observed metal-poor stars but is the result of the different formation process of the bulge. $N$-body simulations show that the first star-forming halos which end up in the bulge are characterized by the highest star-formation rates. These rates enable the formation of rare massive first stars exploding as pair-instability supernovae (PISNe), which wash out the signature of primordial faint supernovae. We demonstrate that the mean $\cfe$ of first stars polluted environments decreases with the increasing contribution of PISNe. We conclude that the dearth of CEMP stars in the Galactic bulge indirectly probes the existence of elusive PISNe, and propose a novel method which exploits this lack to constrain the mass distribution of the first stars.
\end{abstract}


%%%%%%%%%%%%%%%%%%%%%%%%%%%%%%%%%%%%%%%%%%%%%%%%%%

%%%%%%%%%%%%%%%%% BODY OF PAPER %%%%%%%%%%%%%%%%%%

\section{Introduction}
\label{sec:intro}
Today we are surrounded by stars - the Milky Way galaxy alone contains hundreds of billions of stars - but there was a time, billions of years ago, when stars were absent and the Universe was extremely simple. At that time the Universe was mostly neutral and mainly composed by hydrogen and helium produced during the Big Bang Nucleosynthesis.
The first stars played a fundamental role in the evolution of the Universe, as they were responsible for the transition from this very simple and early stage to the more complex one visible today. In fact, the first stars were the sources of the first chemical elements heavier than lithium and of the first hydrogen-ionizing photons. Hence, they initiated the extended processes of {\it reionization} and {\it metal-enrichment}.

Within the standard $\Lambda$CDM model of structure formation, the first stars (referred to as {{\it Population III}} or Pop\,III stars) are predicted to have formed within the first few hundred million years after the Big Bang, corresponding to redshifts of $z \sim 15-20$. 
The primordial birth environment of Pop\,III stars, characterized by lack of heavy elements and dust, may have resulted in a higher Pop\,III characteristic stellar mass with respect to present-day stars \citep[e.g.][]{tan2004formation, hosokawa2011protostellar}
although sub-solar mass stars might also have been able to form \citep[e.g.][]{Greif_2011, Stacy2016}. Since the primordial star formation process is still very poorly understood, we can state that the Initial Mass Function (IMF) of Pop\,III stars is almost completely unknown.
Despite long searches \citep[][]{Beers2005,caffau2013}, zero-metallicity stars have not yet been observed, confirming the hypothesis of a primordial IMF biased towards more massive stars than the present-day IMF \citep[e.g.][]{salvadori2007cosmic,Magg19,rossi2021ultra}. Massive Pop\,III stars explode as supernovae (SNe) polluting the surrounding medium with their chemical products, whose yields depend upon the mass of the progenitor star along with the explosion energy \citep[e.g.][]{HegerWoosley2010,nomoto2004hypernovae}. Hence, even if we cannot directly observe short-lived zero-metallicity stars, we can still catch their long-lived descendants \citep[][]{keller2007, starkenburg2017pristine}. Low-mass (PopII) stars formed in environments enriched by the chemical products of the first stars to the critical metallicity value, $Z>Z_{cr}$, at which normal star formation is expected to proceed \citep[][]{bromm2001fragmentation, schneider2002first}. In this context Stellar Archaeology operates: searching for the chemical signatures of the first stellar generations in the photospheres of old ($> 12$ Gyr) and {{\it metal-poor}} stars that dwell in our Galaxy and its ancient dwarf satellites. 
\\[3pt]

Which components of the Milky Way should be examined to find the oldest living stars? 
The first surveys looking for first star descendants have attempted to select metal-poor stars in the Galactic stellar halo \citep[e.g.][]{Beers_preston, christlieb2003finding}, which is expected to be the most metal-poor component. Typically the iron-abundance\footnote{Throughout this paper we will be using the notation $[A/B] \equiv \log_{10} (N_A/N_B)_\ast - \log_{10} (N_A/N_B)_\odot$, where $N_A$ and $N_B$ refer to the numbers of atoms of elements A and B, respectively.}, [Fe/H], is measured in these surveys and used as a metallicity indicator. The most metal-poor stars are then selected for high-resolution follow-up, often revealing chemically peculiar stars with strong enhancements or deficiencies of particular elements \citep[e.g.][]{christlieb2003finding,caffau2011extremely,Keller2014,bonifacio2015topos,caffau2016topos,Francois2018,aguado2018j0023+,gonzalez2020}.  

The best-known type of chemically interesting stellar object at [Fe/H]$< -2$ is the {{\it carbon-enhanced metal-poor}} (CEMP) class, which has $\cfe>+0.7$ \citep[e.g.][]{Beers2005, Aoki_2007}. This class can be divided into two main populations: (i)~carbon-rich stars that also exhibit an excess in heavy elements formed by the slow neutron-capture processes, having [Ba/Fe]>1 and named CEMP-s stars, and (ii)~carbon-rich stars with no excess of the heavy elements, having [Ba/Fe]<0 and known as CEMP-no stars.
%\footnote{For a detailed taxonomy of CEMP stars see: \citet{Beers2005,Aoki_2007}}. 
The CEMP-s stars are commonly assumed to be chemically enriched by mass transfer from a binary companion star that has gone through the asymptotic giant branch (AGB) phase \citep{abate2015carbon}, and these objects are preferentially found in binary systems \citep[e.g.][]{suda2004he,lucatello2005binary,Starkenburg_2014,hansen2016role_cemps}. 
%On the other hand, CEMP-no stars are not primarily found in binary systems \citep{lucatello2005binary,norris2012most,Starkenburg_2014}\textbf{ - even if \citet{arentsen2019binarity} find a fraction of CEMP-no stars with large absolute carbon abundance to be in binary systems - }hence their C-enhancement is expected to be representative of the interstellar medium (ISM) out of which they formed, which was likely primarily polluted by the first stellar generations \citep[e.g.][]{salvadori15,deBen2016limits}. 
On the other hand, CEMP-no stars are not primarily found in binary systems \citep{lucatello2005binary,norris2012most,Starkenburg_2014,hansen2016role_cempno} and even those that have a binary companion \citep[e.g.][]{arentsen2019binarity} show high values of $\mathrm {^{12}C/^{13}C}$, which implies that the surface composition has not been altered by mass transfer \citep[see][]{Aguado2022,Aguado2023a}. Hence the C-excess in CEMP-no stars is expected to be representative of the interstellar medium (ISM) out of which they formed, which was likely primarily polluted by the first stellar generations \citep[e.g.][]{salvadori15,deBen2016limits}. 

The observed chemical abundance patterns of the most Fe-poor, $\rm [Fe/H] < -4$, CEMP-no stars are indeed consistent with the yields of Pop\,III stars exploding as a {{\it faint supernovae}} and experiencing mixing and fallback \citep[e.g.][]{iwamoto2005first,marassi2014origin}. Because of their low explosion energy, only the outer layers of the Pop\,III progenitor star, which are rich in C and other light elements, can be expelled by faint SNe. On the other hand, the inner part, which is rich in Fe-peak elements, falls back into the center forming a neutron star or a black hole \citep[e.g.][]{HegerWoosley2010}. The increased fraction of CEMP-no stars with decreasing [Fe/H] further supports such a link with Pop\,III star pollution \citep{deBen2016limits}. Furthermore, the high values of $\mathrm {^{12}C/^{13}C}$ recently observed in CEMP-no stars are consistent with an imprint from low-energy Pop~III SNe \citep{Aguado2023a} and rule out the so-called 'spinstars' \citep[e.g.][]{Meynet2006} as main pollutants. CEMP-no stars have been found in a significant fraction in the Galactic halo \citep[][]{Yong_2012,placco2014carbon,carollo2014,lee2017chemical, yoon2018, lee2019chemical} and in the faintest satellites of the Milky Way, the so-called ultra-faint dwarf galaxies (UFDs) \citep[][]{spite2018cemp,Norris_2010,Lai_2011,Gilmore_2013}, which are the oldest galaxies in the Local group \citep[e.g.][]{Simon_2010,gallart2021star}. On the contrary, CEMP-no stars seem to be quite rare in the more luminous classical dwarf spheroidal (dSph) galaxies \citep[e.g.][]{skuladottir2015, yoon2020}, which show more complex and longer star formation histories with respect to UFDs \citep[see][for a global view]{salvadori15}. Ultimately, these observational results confirm that the descendants of the first stars are preferentially found in purely ancient environments. 

Relying on the $\Lambda$CDM model and hierarchical clustering, \citet{white2000first} first predicted the oldest stars to be in the inner part of the Milky Way, i.e. the Galactic bulge.
This idea was confirmed in the following years through different numerical simulations of the Milky Way \citep[][]{diemand2005earth,tumlinson2009chemical,salvadori2010mining,starkenburg2016oldest}. 
More recent simulations have shown this to also hold for different galaxies, including the dwarf satellites of Lyman Break galaxies at redshift $z\approx 6$ \citep{Gelli20}.
Unfortunately, the Galactic bulge is a dusty and overcrowded region, predominantly populated by metal-rich stars, so metal-poor objects are difficult to find \citep[][]{zoccali2008metal,ness2013argos,howes2016embla}.  
The EMBLA Survey (Extremely Metal-poor bulge stars with AAOmega), has been the first in attempting to discover candidate metal-poor stars in this inner region \citep[][]{howes2014gaia,howes2016embla}. Although it successfully identified $\approx 30$ stars with $\feh<-2$, from its observations it is clear that there is a {\it dearth} of CEMP-no stars in this region; in fact, only one CEMP-no star was found, having $\feh=-3.48$ and $\cfe=+0.98$ \citep{Howes2015}. A reason for a lack of CEMP stars in the EMBLA sample could be a selection bias introduced by the SkyMapper photometric selection since photometric colours may have been affected  by the strong CH absorption of CEMP stars with $\rm[C/Fe]>+2$, placing them outside the selected region in the colour-colour diagram. Therefore, the EMBLA Survey could have missed stars with extremely large C enhancements but stars with mild C enhancements should still have been found.

More recently, the Pristine Inner Galaxy Survey \citep[PIGS][]{arentsen2020pristine} targeted the Galactic bulge with low/intermediate resolution spectroscopy (R $\approx 1300$ at 3700-5500 \AA, and $R\approx 11000$ at 8400-8800 \AA), collecting 1900 stars with $\feh < -2.0$, which is currently the largest sample of confirmed very metal-poor stars in the inner Galaxy. Since s-process abundance measurements for this sample are unavailable, a different CEMP classification was applied based on the absolute carbon abundance\footnote{Elemental abundances can also be referred to as an “absolute” scale, relative to the number of hydrogen atoms, defined as $A(C) \equiv \log_{10} (N_C/N_H)_\ast + 12$.}, A(C), and the \feh of the stars \citep[][]{yoon2016observational}. According to \citet{bonifacio2015topos} \citep[see also][]{spite2013carbon} CEMP stars can be divided in two main groups:
(i) the high-carbon band, A(C) > 7.4, largely containing CEMP-s stars that typically have higher iron-abundance; (ii) the low-carbon band, A(C) < 7.4, predominantly containing CEMP-no stars at lower [Fe/H]. Using this classification while correcting for evolutionary effects in Red Giant Branch (RGB) stars \citep[e.g.][]{placco2014carbon}, the PIGS survey identified 24 new CEMP-no candidate stars in the Galactic bulge. Still, the overall fraction of CEMP-no stars obtained by PIGS is only $\lesssim 6\%$ at [Fe/H]$<-2$, i.e. much lower than what is found in the Galactic halo ($\approx 20\%$, see \citealt{arentsen2021pristine}). Furthermore, $\simeq96\%$ of their CEMP-no candidates have only a moderate C-enhancement, $ +0.7 < \cfe < +1.2$, with only one CEMP-no star at $\feh \simeq -3.5$ that shows a high $\cfe \simeq +2.2$, while similar values are more frequently observed both in the Galactic halo and in UFDs \citep[e.g. Fig.~1 from][]{salvadori15}. As explained in \citet{arentsen2021pristine} and \citet{arentsen2022}, the selection of metal-poor stars in PIGS could be significantly affected by biases in the photometry against CEMP stars with strong CH lines, i.e. those with very
high carbon abundances and/or cooler temperatures. However, stars with $\feh < -3.0$ are not expected to be substantially biased, except for the coolest ($T_{\mathrm{eff}} < 4750\, \mathrm{K}$) and most carbon-rich ($\cfe > +2.0$) stars. For stars with $\feh < -2.0$ and $< -2.5$, slightly warmer and less carbon-rich stars are also affected, but still most of the CEMP-no stars are expected to stay within their selection. Therefore they expect that the CEMP-no fraction is less impacted by a photometric bias compared to that of CEMP-s stars.

These observational results raise several questions: Why is there an apparent dearth of CEMP-no stars in the Galactic bulge? And why are CEMP-no stars with high \cfe so rare in this ancient environment? Is it entirely due to a bias in the selection of metal-poor stars or is there a process that reduces the CEMP-no fraction in the Galactic bulge? The aim of this paper is to answer these questions, and it is structured as follows. In Section \ref{sec:obs} we will carry out a preliminary analysis on the observational data of metal-poor stars in the different regions of the Local Group in order to understand whether the dearth of C-enhanced stars in the Galactic bulge is due to a statistical effect. In Section \ref{sec:model} we will illustrate the cosmological model used to follow the evolution of the Galaxy that combines a 
\textit{N}-body simulation, following the hierarchical assembly of a Milky Way (MW)-like galaxy, with a semi-analytical model that follows the evolution of baryons. The results obtained from this model and the additional analytical calculations performed will be described in Sec. \ref{sec:results}. Finally, in Sec.~\ref{sec:concl} we will draw the conclusions of our analysis with the associated implications, and we will list the prospects for future works.

\section{Observational data analysis}
\label{sec:obs}

To understand why CEMP-no stars are less common in the Galactic bulge with respect to other ancient environments of the Local Group, we should first analyze if this dearth can be just a statistical effect. Indeed, since CEMP-no stars are more common at decreasing $\feh$, the probability to discover them in environments dominated by metal-rich stars, such as the Galactic bulge, can be intrinsically very low \citep{salvadori15}. 

\begin{figure}
    \centering
    \includegraphics[width=.465\textwidth]{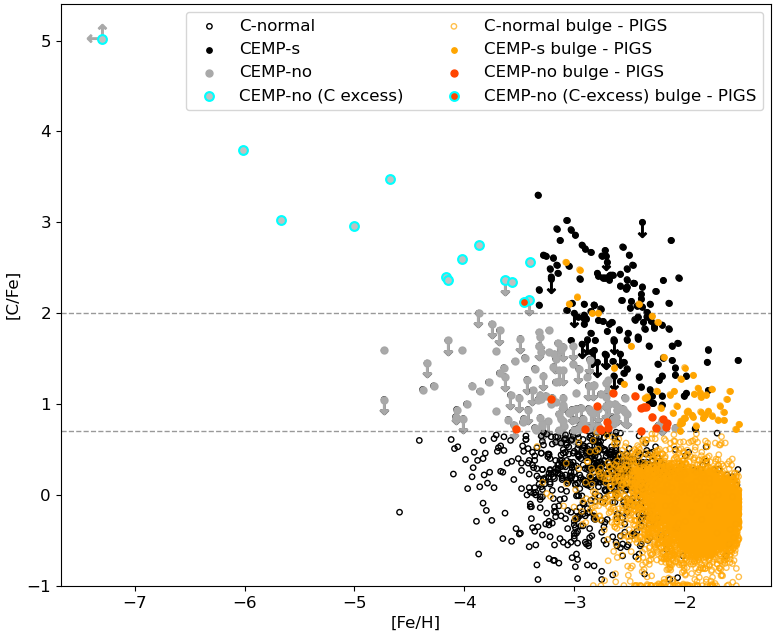} 
    \includegraphics[width=.48\textwidth]{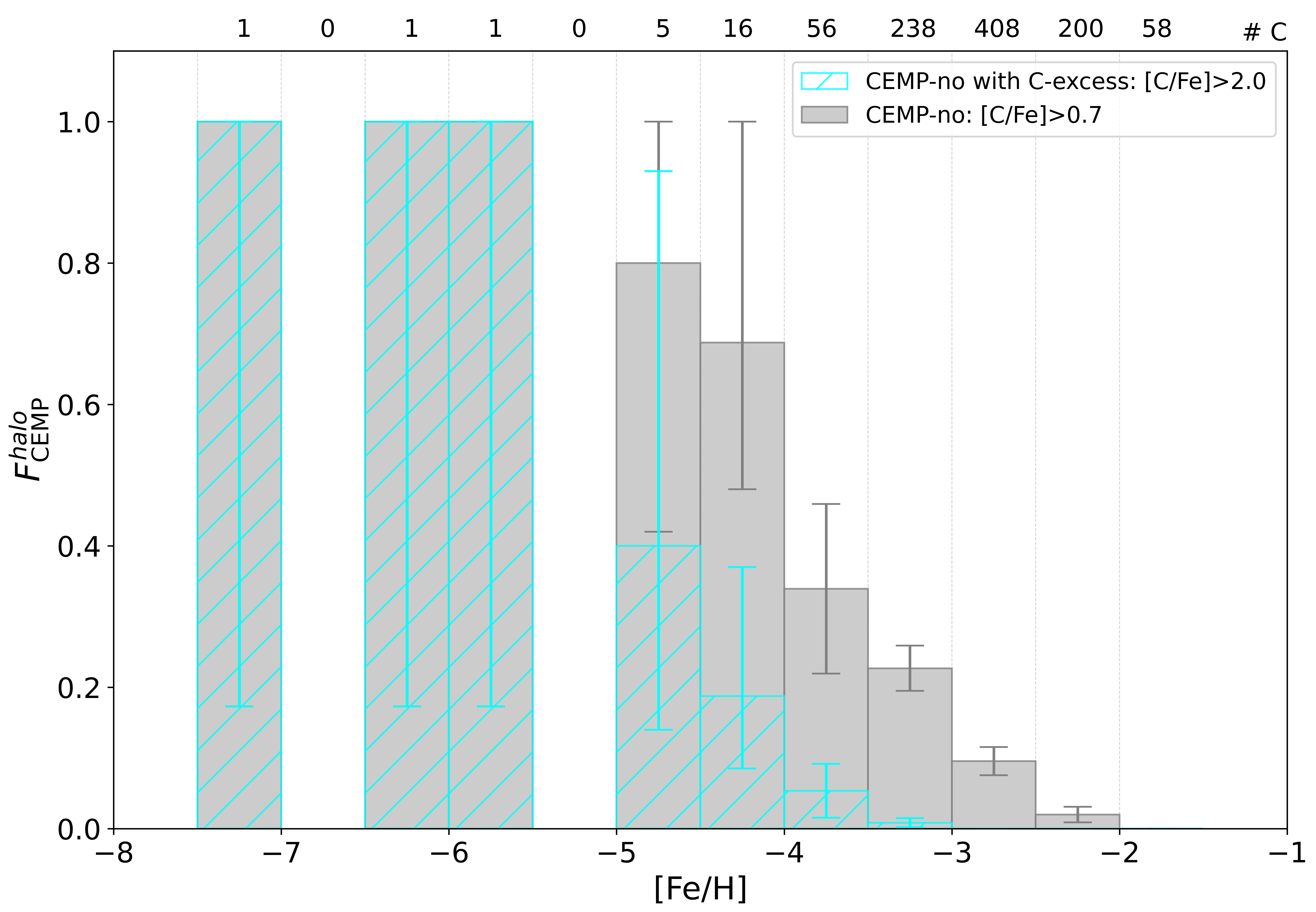} 
    \caption{\emph{Top panel}: Measured \cfe vs \feh for stars in the Galactic halo (black and grey points; {\tt{JINAbase}}) and the bulge PIGS sample (orange and red points; \citealt{arentsen2021pristine}). \emph{Bottom panel:} Fraction of CEMP-no stars in the Galactic halo in different $\ch{[Fe/H]}$ ranges: gray includes all CEMP-no stars, while cyan is only CEMP-no stars with carbon-excess, $\cfe > +2.0$. Error bars are Poissonian errors derived from \citet{gehrels1986confidence}. The number of stars in each bin, $N_{*}(\feh)$, is listed on top.}
    \label{fig:fcemp_halo}
\end{figure}

\begin{figure*}
\begin{multicols}{2}
\includegraphics[width=.48\textwidth]{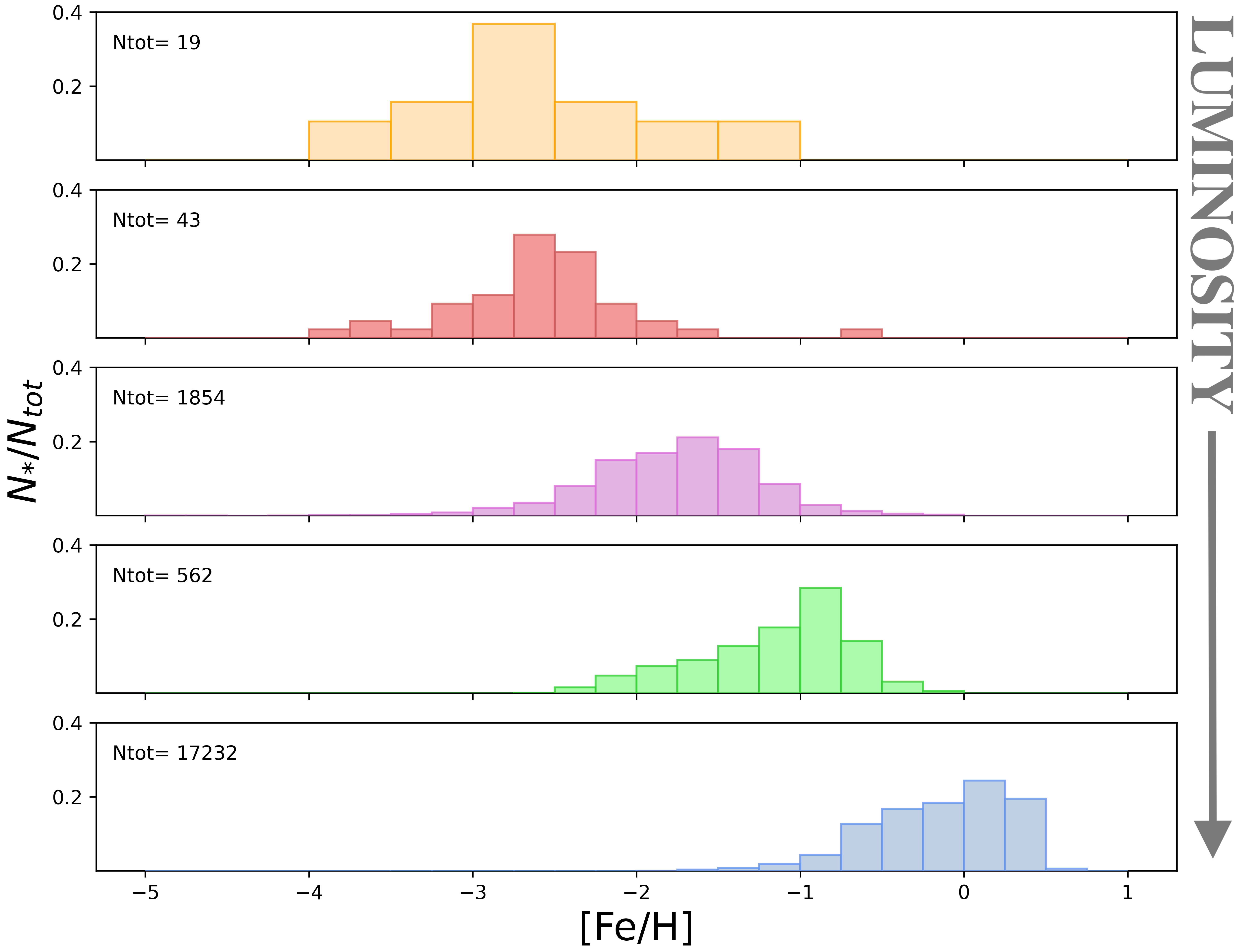}
\par
\includegraphics[width=.48\textwidth]{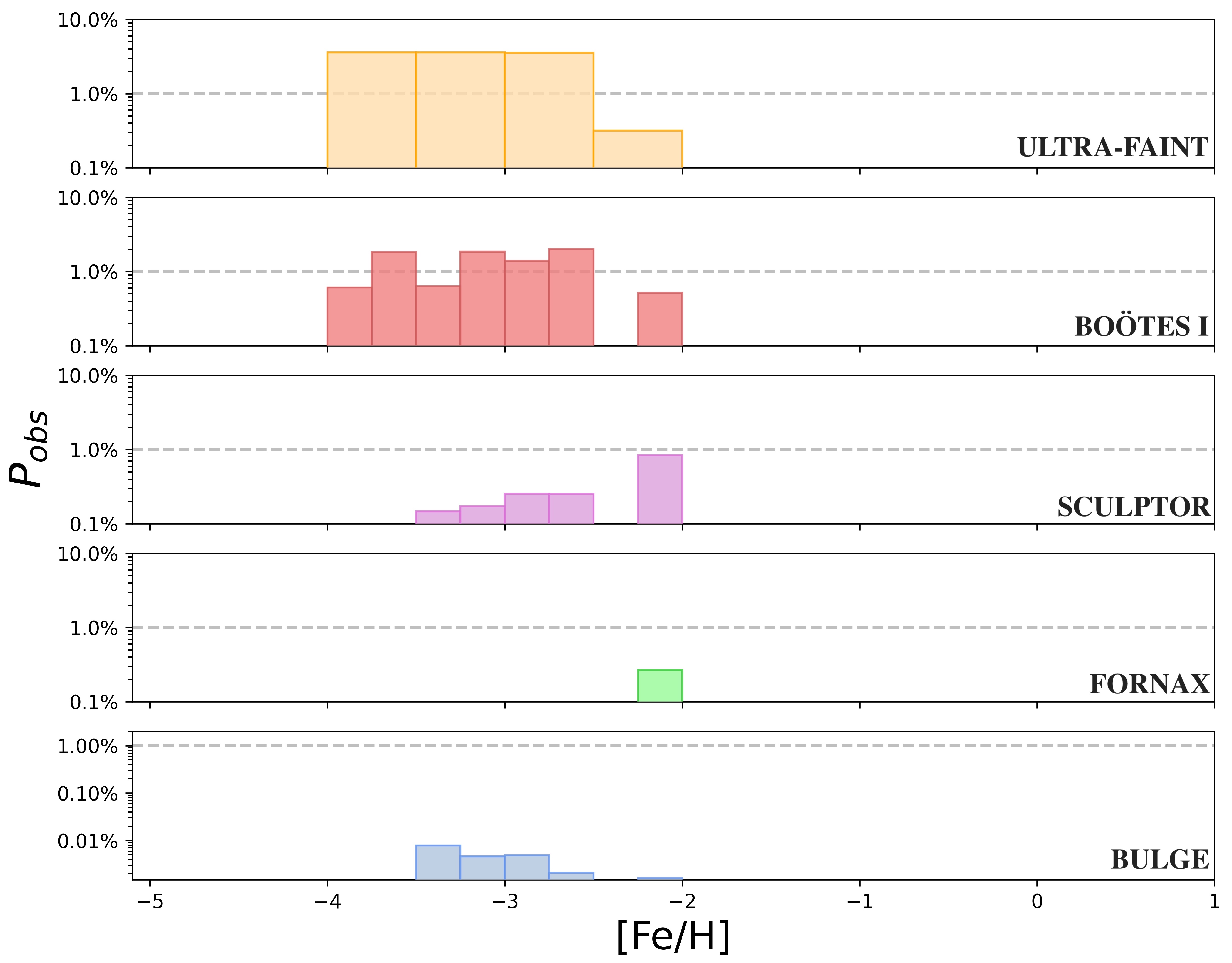}
\par
\end{multicols}
\caption{The MDF (left panel) and the probability of observing a CEMP star (right panel) for the different environments displayed in order of increasing luminosity (mass): ultra-faint galaxies (orange), Boötes (red), Sculptor (pink), Fornax (green) and the Galactic bulge (blue).}
\label{fig:ordered}
\end{figure*} 

\subsection{The fraction of CEMP stars} \label{sec:CEMPfrac}
 Henceforth we will simply speak of CEMP stars when referring to the subclass of CEMP-no stars, i.e to stars with $\feh<-2$, $\cfe>+0.7$ within the low-carbon band, A(C) < 7.4. The upper panel of Figure \ref{fig:fcemp_halo} shows the measured \cfe vs \feh for a sample of 984 halo stars with carbon measurements from the {\tt{JINAbase}}{\footnote{\url{https://jinabase.pythonanywhere.com/}}\citep{abohalima2018jinabase}}. The abundance and stellar parameter data collected there are based on high-resolution spectroscopic studies ($R = \lambda/\Delta\lambda \gtrsim 15\,000$, with the majority having $R = 30\,000 - 40\,000$) found in the literature. In the same plot we also show the measurements for the sample of bulge stars (Arentsen private comm.) as selected in \citet{arentsen2021pristine} in their Fig.~9. In detail they select measurements from a FERRE analysis with the following cuts on gravity, temperature and \cfe uncertainty: $\mathrm{log}g<3.5$, $\mathrm{4600\, K < T_{eff} < 5500\, K}$ and $\epsilon_{\cfe} < 0.5$. 
 As stated in Sec.~\ref{sec:intro}, for bulge stars there are not available measurements of the s-process element Ba. Thus, the A(C) classification is used to distinguish among CEMP-no and CEMP-s candidates. Although many halo stars have Ba measurements, here we adopt the same definition to separate CEMP-no and CEMP-s stars to have a self-consistent classification. As explained in detail in \citet{yoon2016observational}, despite the differences in the two criteria, the clear distinction in A(C) between CEMP-s and the CEMP-no stars in their halo sample appears to be as successful, and likely more astrophysically fundamental, for the separation of these sub-classes as the previously recommended criterion based on [Ba/Fe] abundance ratios. 
 Fig.~\ref{fig:fcemp_halo} (upper panel) clearly shows how the \cfe values in Galactic halo stars are higher at $\feh < -3$ compared to stars in the bulge, and how in general CEMP-no stars are more frequent in the halo, although we cannot exclude that this can be due to a statistical effect given the rarity of extremely metal-poor stars in the Galactic bulge (see also Sec.~\ref{sec:cemp_hidden}). To quantify this, we compute the fraction of CEMP-no stars, defined as the ratio of CEMP-no stars over the total number of stars at a given $\feh$:

\begin{equation}
    F_{\mathrm{CEMP}}(\feh) = \frac{N_{\mathrm{CEMP}}(\feh)}{N_{*}(\feh)}.
\end{equation}\\

The Galactic halo is currently the best observed ancient environment of the Local Group and for which we have the largest statistics regarding the chemical properties of the metal-poor stellar population \citep[e.g.][]{yong2012bmost,bonifacio2021topos}. Therefore we compute $F^{halo}_{\mathrm{CEMP}}$ for the sample of halo stars in Fig.~\ref{fig:fcemp_halo}. %\agu{It actually includes observations until 2017, not included.}. 
The results are shown as grey histogram in the bottom panel of Fig.~\ref{fig:fcemp_halo} where we have also computed $F^{halo}_{\mathrm{CEMP}}$ for CEMP stars with high carbon-excess, $\cfe>+2.0$ (cyan histogram).
%, where for comparison we also display $F_{\mathrm{CEMP}}$ obtained from high-resolution measurements for $\rm \sim 120$ halo stars at [Fe/H]$< -3$ \citep[see][and references therein]{deBen2016limits}. 
As we can see, $F_{\mathrm{CEMP}}$ increases as $\feh$ decreases and reaches the value of $\approx 1$ for $\feh \leq -5$. Because of the large number of CEMP-no stars at $\feh \leq -4 $ discovered during the last 10 years \citep[see e.g.,][and references therein]{Yong_2012,norris2012most,Keller2014,bonifacio2015topos,agu17II,Dacosta2019,nordlander2019,li2022}, we can now derive $F_{\mathrm{CEMP}}$ at different [Fe/H] values, equally spaced every 0.5 dex. Still, we should remind that these values suffer various systematic uncertainties. On one hand, as suggested in \citet{arentsen2022}, high-resolution spectroscopic compilations of very metal-poor stars ($\feh < -2.0$) might be biased towards (very) carbon-rich objects at the ‘high’ metallicity end ($\feh > -3.0$) due to their follow-up strategies. This leads to an overestimate of the CEMP fraction in this metallicity range. Furthermore, $F^{halo}_{\mathrm{CEMP}}$ might increase when correcting the carbon measurements to account for the internal depletion of carbon occurring in RGB stars \citep{placco2014carbon}. On the other hand, $F^{halo}_{\mathrm{CEMP}}$ might decrease when accounting for non-local thermodynamic equilibrium (non-LTE) effects, which can drastically lower the measured [C/Fe] value \citep{amarsi2019carbon}. Finally, the selection function of the data might suffer some bias in either direction.
Since the main two effects compensate and non-LTE corrections are only available for a handful of stars \citep{amarsi2019carbon}, we will consider the derived $F^{halo}_{\mathrm{CEMP}}$ as the reference fraction of Galactic halo stars at different iron abundances. It is also worth mentioning some work suggesting that the frequencies of CEMP varies with location in the Galactic halo \citep[][]{yoon2018,lee2018}.
Despite the mentioned caveats, as an initial hypothesis, we will assume that the fraction of CEMP-no stars is {\it the same in all environments} and equal to that of the Galactic halo, $F_{\mathrm{CEMP}} \equiv F^{halo}_{\mathrm{CEMP}}$.
\\[2pt]
\subsection{Ancient environments in the Local Group}
To understand whether CEMP-no stars might simply be hidden in the Galactic bulge, we need to derive the probability to catch CEMP-no stars while blindly observing the bulge and then compare to other ancient environments in the Local Group \citep{salvadori15}. To this end we analyzed a number of environments with {\it increasing} stellar mass (luminosity) and correspondingly an {\it increasing} average stellar metallicity. In particular we considered (see Table~\ref{tab:envs} for details): the least luminous UFDs,
%\hspace{0.5cm}$ L\leq 10^4 L_\odot$, $\langle\feh\rangle\lesssim-2.2$\;
the most luminous UFD {\it Bootes I},
%\hspace{0.5cm} $ L \approx 10^{4.5} L_\odot$, $\langle\feh\rangle\sim-2.1$;
the dSph galaxy {\it Sculptor},
%\hspace{0.5cm}$ L \approx 10^{6.34} L_\odot$, $\langle\feh\rangle\sim-1.8$, 
 the dSph galaxy {\it Fornax} and  
%\hspace{0.5cm}$ L \approx 10^{7.25} L_\odot$, $\langle\feh\rangle\sim-1.0$
the Galactic {\it bulge}. 
%\hspace{0.5cm}$ L \approx 10^{10} L_\odot$, $\langle\feh\rangle\sim0.0$.

\begin{table}
\caption{Luminosities and average stellar metallicities of the Local Group environments taken into account in this work.}
\resizebox{\columnwidth}{!}{
\begin{tabular}{|c|c|c|c|c|c|}
\hline
\multicolumn{1}{|l|}{} &
  \begin{tabular}[c]{@{}c@{}}least luminous\\ UFDs\end{tabular} &
  Bootes I &
  Sculptor &
  Fornax &
 \begin{tabular}[c]{@{}c@{}}Galactic\\ bulge\end{tabular} \\ \hline
$L_{*}$ &
  $\leq 10^4 L_\odot$ &
  $10^{4.5} L_\odot$ &
  $10^{6.34}L_\odot$ &
  $10^{7.25} L_\odot$ &
  $10^{10} L_\odot$ \\ \hline
$\langle\feh\rangle$ &
  $\lesssim-2.2$ &
  $-2.1$ &
  $-1.8$ &
  $-1.0$ &
  $0.0$ \\ \hline
\end{tabular}}

\label{tab:envs}
\end{table}

The carbon measurements in the least luminous UFDs are from high-resolution spectroscopic studies ({\it Segue 1}: \citet{Norris_2010}; \citet{Frebel_2014}. {\it Pisces II}: \citet{refId0}. {\it Ursa Major II and Coma Berenice}: \citet{Frebel_2009}. {\it Leo IV}: \citet{Simon_2010}). Only few stars have been observed in each UFD because these galaxies are faint and distant so that only few RGB stars are available for spectroscopic observations. For this reason we consider stars belonging to the faintest UFDs all together. 
Stars in {\it Boötes I}, the most luminous UFD, are studied with low-resolution spectroscopy \citep[][]{Lai_2011,Norris_2010} with the only exception of seven stars \citep{Gilmore_2013}. 
In the {\it Sculptor} dSph galaxy, many carbon measurements are available from both low- \citep{Kirby_2015} and high-resolution spectroscopic studies \citep[][]{Frebel2010,tafel_2010,Starkenburg_2014,Simon_2015,Jablonka_2015,skuladottir2015}. 
Finally, for the {\it Fornax} dSph we include the low-resolution measurements from \citet{Kirby_2015}.

For the Galactic bulge, we use data from the {\tt JINAbase} and those provided by the high-resolution, large multi-object APOGEE spectroscopic survey \citep[Data Release 16,][]{jonsson2020apogee}.
%\ag{We already have APOGEE DR17, shall we use it?}

\subsection{Are CEMP stars hidden in the Galactic bulge?}
\label{sec:cemp_hidden}

For a ``blind'' survey that does not pre-select the most metal-poor stars, the ability to find CEMP stars is naturally limited by the number of stars that exist at each $\feh$, i.e. the Metallicity Distribution Function \citep[MDF,][]{salvadori15}. In Fig.~\ref{fig:ordered} (left), we show the normalized MDFs for the different environments studied. %, in order of increasing brightness from the least luminous dwarf galaxies to the bulge. ÁS: We shouldn't repeat what is in the caption.
Note that the total number of stars observed in these environments strongly increases with galaxy luminosity: only 19 stars constitute the MDF in the faintest UFDs, while for the Galactic bulge we have $>17,000$~stars. 

We then ask ourselves: if we assume that the fraction of CEMP stars is ``Universal'' and equal to that of the Galactic halo, $F^{halo}_{\mathrm{CEMP}}$, what is the joint probability, $P_\text{obs}$, to observe a star that has a given [Fe/H] value, which is also carbon-enhanced? This is done by combining two independent functions: the halo fraction of CEMP stars and the normalized MDF derived for each environment:
\begin{equation}
P_{\mathrm {obs}}(\rm[Fe/H]) = F^\textsl{halo}_{\mathrm{CEMP}} (\rm[Fe/H]) \times \frac{N_*(\rm [Fe/H])}{N_{tot}}.
\end{equation}

In the right panel of Fig. \ref{fig:ordered}, we show ${P_{\rm obs}}$ for the different environments studied, 
and we note a clear trend: the brighter is the galaxy, and thus more shifted towards higher $\feh$ values, the lower is the overall probability to observe a CEMP star. 
In other words, in the most massive galaxies, stars in the low-$\feh$ tail become almost ``invisible'' with respect to the metal-rich stars since they are much rarer. 
This is especially evident in the bulge since the peak of the resulting probability ($P_{\mathrm {obs}}<0.01\%$) is two orders of magnitude lower than the values obtained in UFDs ($P_{\mathrm{obs}}\sim 3\%$), where more CEMP stars have actually been discovered. This peak of $P_{\mathrm {obs}}$ in the bulge, at $\feh\simeq -3.5$, is almost flat for $-3.5<\feh<-2.0$, and drops significantly at $\feh<-3.5$. %Furthermore, the peak of $P_{\mathrm {obs}}$ in the bulge is shifted towards higher $\feh$ values with respect to all dwarf galaxies.

In conclusion, the scarcity of stars at low-$\feh$, makes it very difficult to search for CEMP stars in luminous environments, such as the Galactic bulge, that are dominated by metal-rich stars. The dearth of CEMP stars in the Galactic bulge can therefore partially be ascribed to this reason. However, neither EMBLA or PIGS are ``blind'' surveys since they specifically search for the most metal-poor stars in the Galactic bulge. PIGS, in particular, obtained a sample of 1900 stars with $\feh<-2$. Thus, if we hypothesize that the CEMP fraction within the bulge is equal to the one we obtained for the Galactic halo, namely $F^{bulge}_{\mathrm {CEMP}}(\feh<-2)=F^{halo}_{\mathrm {CEMP}}(\feh<-2)\approx 24\%$, then we can estimate that $\sim450$ CEMP stars should have been observed by PIGS. Instead, $<62$ CEMP(-no) candidates have been identified so far \citep{arentsen2021pristine}. In agreement with the PIGS papers, we thus conclude that the dearth of CEMP stars in the Galactic bulge cannot be only a consequence of a statistical effect in this metal-rich environment. It it is also interesting that we derive a peak probability of observing CEMP stars in the Galactic bulge at $\feh\simeq-3.5$, i.e. at the metallicity of the only CEMP star with high \cfe observed by \citet{arentsen2021pristine}.

\section{Cosmological model description}
\label{sec:model}

{In this Section we will briefly recap the cosmological model used to identify the first star-forming halos whose stars are currently located in the Galactic bulge and to investigate their properties. The model combines a $N$-body simulation that follows the hierarchical assembly of a Milky Way (MW) - like galaxy \citep[used also in][]{salvadori2010mining, pacucci2017gravitational} with a semi-analytical model \citep[{\tt{GAMETE}},][]{salvadori2007cosmic,salvadori15} that follows the evolution of baryons, from the formation of the first stars to the present day. The semi-analytical model allows us to follow the star formation and metal enrichment history of the Milky Way from early times ($z\sim20$) until now ($z=0$) and thus to link the chemical abundances of present-day stars with the properties of the first stellar generations.

\label{subsec:nbody}
\begin{itemize}
\item{\bf N-body simulation}
The $N$-body simulation used to study the hierarchical formation of the MW has a low-resolution region corresponding to a sphere of radius $10h^{-1}$\,Mpc \citep[see][]{scannapieco2006spatial}. The region of the highest resolution is a sphere of radius $1h^{-1}$\,Mpc, i.e. four times the virial radius of the MW at z = 0 ($r_{vir} = 239$ kpc). 
A low-resolution simulation including gas physics and star formation has been used to confirm that the initial conditions will lead to a disk galaxy like the Milky Way. 
The system consists on about $10^{6}$ dark matter (DM) particles within $r_{vir}$ with masses of $7.8 \times 10^{5} M_{\odot}$; its virial mass and radius are respectively $M_{vir} = 7.7 \times 10^{11} M_{\odot}$ and $r_{vir} = 239$ kpc, roughly consistent with the observational estimates of the MW, $M_{vir}\approx 10^{12} M_{\odot}$ \citep[e.g.][]{McMillam}. The softening length is 540\,kpc. The simulation data is output every 22 Myr between $z = 8 - 17$ and every 110\,Myr for $z < 8$. At each output, the virialized DM halos have been identified using a {\it friend-of-friend} group finder with a {\it linking parameter} $b = 0.15$, and a threshold number of particles constituting virialized halos equal to 50. The $N$-body simulation enables to reconstruct a hierarchical history of the MW that proceeds through the consecutive merging of DM halos maintaining the information about the spatial distribution of DM particles belonging to them.
\\[2pt]

\item {\bf Star formation}
At the initial redshift of the simulation, $z = 20$, gas in DM halos is assumed to have a primordial composition, and only objects with virial temperatures $T_{vir}\geq 10^4$ K form stars. This choice, which is equivalent to assuming that the star formation activity is rapidly quenched in mini-halos, is dictated by the limited DM resolution of the $N$-body simulation \citep{graziani2015galaxy}.
At each time-step, stars are assumed to form in a single burst, proportional to the available cold gas mass, $M_{gas}$. The constant of proportionality is a redshift-dependent star-formation efficiency $\rm f_{*}(z)=\epsilon_* \; \frac{\Delta t(z)}{t_{ff}(z)}$, where $\rm t_{ff}$ is the free-fall time, $\rm \Delta t(z)$ the $N$-body time-step and $\epsilon_{*}$ a free parameter of the model, physically corresponding to a {\it “local” star formation efficiency}.
\\[2pt]
\item {\bf Pop\,III stars}
The model simply assumes that the Pop\,III IMF is a delta function centered either on the average mass value of the pair-instability SNe \citep[PISNe;][]{salvadori2010mining} or faint SNe \citep{pacucci2017gravitational}, i.e:
    \begin{equation}
    \label{eq:pisn_imf}
    \Phi(m) = \frac{dN}{dm} \propto \delta(m_{\rm pop\,III}).
    \end{equation}
To give an idea of the amount of chemical elements released by these different Pop\,III stars, an ``average'' PISN ($\rm m_{Pop\,III}=200\,M_{\odot}$) yields of metals, iron, and carbon respectively is equal to: $Y= M_Z/M_{\rm pop\,III} = 0.45$, $Y_{\ch{Fe}} = 0.022$ and $Y_{\ch{C}} = 0.02$ \citep{heger2002nucleosynthetic}. Conversely, an average faint SN ($\rm m_{*} =25\,M_{\odot}$) gives $Y \sim 0.1$, $Y_{\ch{Fe}} = 4\times10^{-7}$ and $Y_{\ch{C}} = 9.98\times10^{-3}$ \citep[][]{iwamoto2005first,marassi2014origin}.\\    

\item {\bf Pop\,III-to-Pop\,II transition}
Following the critical metallicity scenario \citep[e.g.][]{bromm2001fragmentation,schneider2002first} we assume that the IMF of newly formed stars depends upon the initial metallicity of the star-forming clouds. Therefore, a star forming halo with a gas metallicity $Z \leq Z_{cr}$ will form Pop\,III stars, otherwise, if $Z > Z_{cr}$, it will form Pop\,II/I stars. Exploiting the data driven constraints of \citet{deBen2016limits}, we set $Z_{cr} = 10^{-4.5}Z_{\odot}$. Normal Pop\,II/I stars are assumed to have masses in the range $[0.1, 100]\,M_{\odot}$ and to form according to a standard Larson IMF \citep{larson1998early}, which peaks at the characteristic mass $\rm m_{ch} = 0.35M_{\odot}$ and rapidly declines with a Salpeter-like shape towards larger masses \citep{salpeter1955luminosity}. \\ 

\item {\bf Metal enrichment}
The newly formed stars are assumed to instantaneously evolve at the following snapshot of the simulation (Instantaneous Recycling Approximation, IRA) since the time elapsed between two neighbouring steps is larger than the lifetime of the least massive stars evolving as SNe. For Pop\,III stars exploding as PISN we assume the yields of \citet{heger2002nucleosynthetic}, while for primordial faint SN we assume those of \citet{iwamoto2005first}. For Pop II/I stars we adopt the yields from \citet{woosley1995evolution}.
To not overestimate the contribution of carbon due to AGB stars that also produce slow neutron-capture process elements and, even if not in binary systems, can lead to the formation of "moderate" CEMP-s stars \citep{rossi2023}, we only considered the chemical products of Pop\,II/I that evolve as core-collapse SNe in short timescales ($3-30$ Myr). 
After SN explosions the newly produced/injected metals are assumed to be instantaneously and homogeneously mixed within the Inter Stellar Medium (ISM) eventually reaching the Inter Galactic Medium (IGM) via supernova driven outflows.
\\[2pt]

\item {\bf Gas and metals dispersal}
Supernovae may release a high amount of energy, which may overcome the binding energy of the hosting halo leading to a partial gas and metals removal from the galaxy.
The mass of gas ejected into the IGM, $\rm M^{ej}_{gas}$, depends on the balance between the escape velocity of the halo, $v_{esc}$, and the kinetic energy released during the explosion, namely:
\begin{equation}
\rm M^{ej}_{gas}=(2E_{SN})/v^2_{esc}
\label{eq:m_ej}
\end{equation}
where $\rm E_{SN} = \epsilon_w\,N_{SN}\,\langle E_{SN}\rangle$, with $\rm N_{SN}$ number of SN explosions and $\rm \langle E_{SN}\rangle$ the average explosion energy. For a typical PISN ${\rm \langle E^{PISN}_{200M_{\odot}}\rangle\sim 2.7\times 10^{52}}$ erg \citep{heger2002nucleosynthetic}, while a typical faint SN provides $\rm \langle E^{faint}_{25M_{\odot}}\rangle\sim 0.7\times 10^{51}$ erg \citep{iwamoto2005first,marassi2014origin}, which is lower than the energy released by a normal   $25\,M_{\odot}$ core-collapse SN, $\rm E^{cc}_{25M_{\odot}}\approx10^{51}$ erg.

The quantity $\epsilon_w$, representing the second free parameter of the model, is the {\it wind efficiency}, that is the fraction of the explosion energy converted into kinetic form. 
\\[2pt]

\item {\bf Chemical evolution}
Due to mechanical feedback, the mass of gas and metals in a halo can decrease substantially. At each simulation time-step the gas mass reservoir in each halo, $\rm M_{gas}$, is updated with respect to the initial gas mass, $\rm M^{in}_{gas}$, to account for the mass of stars locked into newly formed stars, $\rm M_{*}$, and the gas mass ejected out of the halo, $\rm M^{ej}_{gas}$:
\begin{equation}
\label{eq:m_gas}
    \rm M_{gas} = M^{in}_{gas} - (1+R) M_* - M^{ej}_{gas}.
\end{equation}
where $R$ is the returned fraction, which is equal to 1 only for PISN and lower than unity otherwise.
Similarly, the mass of metals, $\rm M_{Z}$, in each hosting halo is updated as follow:
\begin{equation}
    \rm M_{Z} = {M^{in}_{Z}+Y\,M_{*} - Z^{in}_{ISM}M_*- Z^{in}_{ISM}M^{ej}_{gas}},
\label{fig:m_Z}
\end{equation}
where $\rm M^{in}_{Z}$ is the initial mass of metals, $\rm Y_{Z}$ the metal yield, and $\rm Z^{in}_{ISM}$ the initial metallicity of the ISM.
\\[2pt]

\item {\bf Model calibration}
To set the best values of the two model free parameters, $\epsilon_*$ and $\epsilon_w$, the observed properties of the MW have been used as a benchmark. In particular, the results of the simulations at redshift $z = 0$ have been compared with the gas/stellar mass and metallicity of the MW, the baryon-to-dark matter ratio, and the metallicity of high-velocity clouds \citep[see][for more details]{salvadori2007cosmic,salvadori2010mining}.
\end {itemize}

\section{Simulation analysis and results}
\label{sec:results}

\subsection{Extreme cases of Pop\,III enrichment} 
\label{sec:extreme}
To understand how much the properties of Pop\,III stars can affect the present-day distribution of CEMP stars, we first analyse the $z=0$ outputs of the simulation for two extreme cases: (i)~when all Pop\,III stars explode as an average PISN of $\rm 200\,M_{\odot}$, or (ii)~when they all explode as an average faint SN of $\rm 25\,M_{\odot}$ (see Sec.~\ref{sec:model}). In the first case we find that CEMP stars are never produced by the model, the carbon-to-iron value of present-day stars being $\cfe \lesssim 0.2$. %{\bf Please, Giulia, double check this!} 
This is perfectly consistent with the low [C/Fe] value of an ISM imprinted by massive $> 160\,M_{\odot}$ Pop\,III stars evolving as PISN \citep[e.g.][left panel of Fig.~2]{salvadori2019probing}.

Conversely, CEMP stars are formed when all Pop\,III stars are assumed to explode as $25\,M_{\odot}$ faint SNe and they can have huge C-excess, [C/Fe]$\geq +4$, at the lowest [Fe/H]$<-4$. This result is in line with the idea that the most iron-poor CEMP(-no)
stars are indeed the descendants of this kind of zero-metallicity SNe (Sec.~\ref{sec:intro}). Under this assumption we can also compute the present-day distribution of CEMP stars which predicts that at $\feh<-2$ the innermost region hosts the largest mass fraction of CEMP stars. This remains true even if we only consider CEMP stars with $\cfe > +2.0$, see Fig.~\ref{fig:CEMP_distr}. 

Although we cannot completely rule out a statistical effect (see Sec.~\ref{sec:cemp_hidden}), these results are at odds with observations, which actually suggest a \textit{lower} fraction of CEMP stars in the bulge than in the outer regions (Sec.~\ref{sec:CEMPfrac}). Therefore it is clear that to model the early chemical enrichment in the first star forming halos it is necessary to consider a more realistic Pop\,III IMF, which accounts for the contribution of both faint SNe and PISNe.
\begin{figure}
   \centering
    \includegraphics[width=0.5\textwidth]{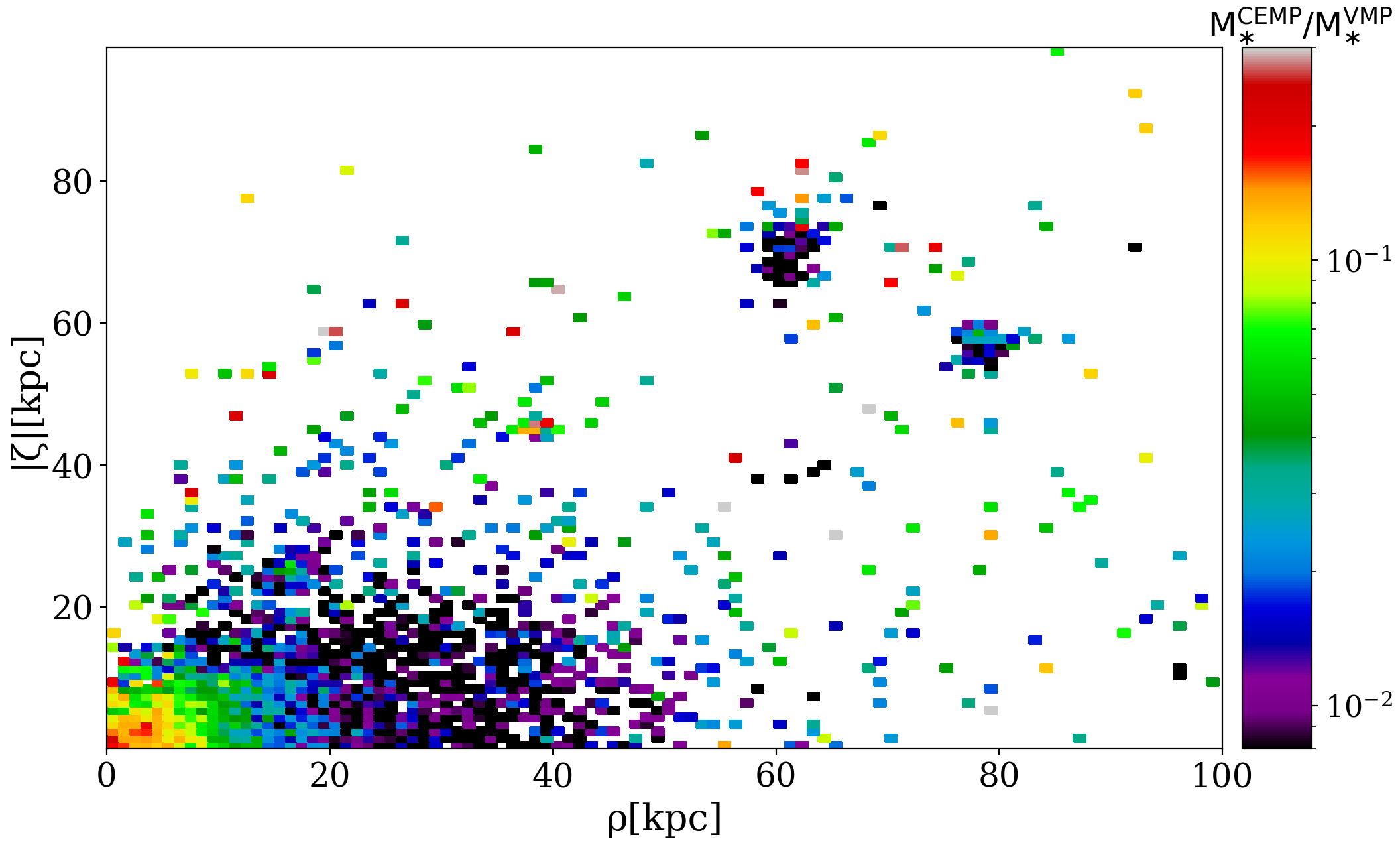}\hfill%
    \caption{Present-day mass fraction of CEMP stars with $\cfe> +2 $, relative to all stars at $\feh < -2 $, contained in an annulus of radial width within 1 kpc (see colormap) plotted in the $\rho - \zeta$ cylindrical coordinate plane of our simulated MW-analogue galaxy. The results have been obtained from 
    our semi-analytical model by assuming that {\it all} Pop\,III stars have $\rm m_{pop\,III}=25M_{\odot}$ and explode as faint SNe.   
    }
    \label{fig:CEMP_distr}
\end{figure}

\begin{figure}
\begin{centering}
\includegraphics[width=.75\linewidth]{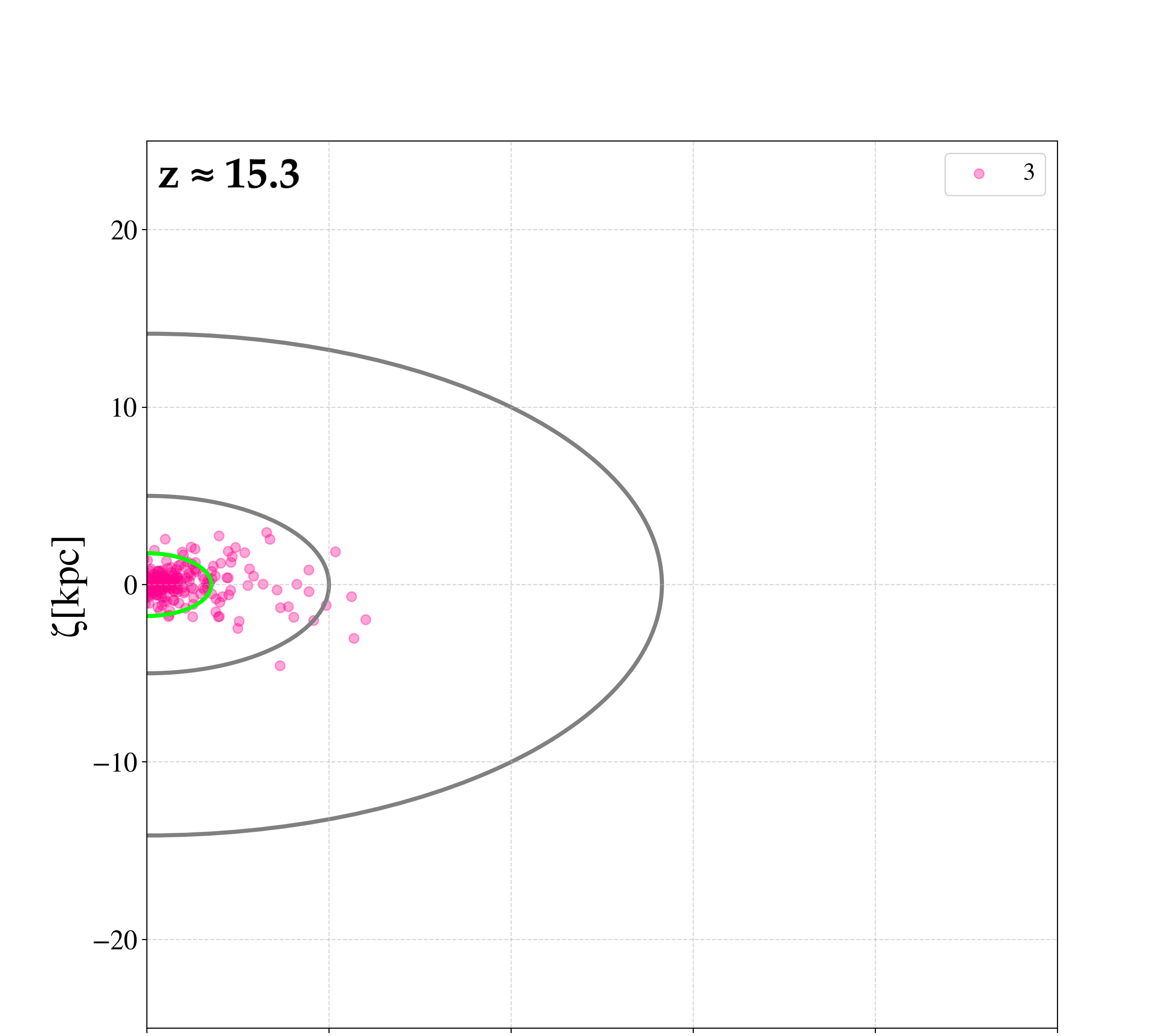} \par \vspace{-22pt}
\includegraphics[width=.75\linewidth]{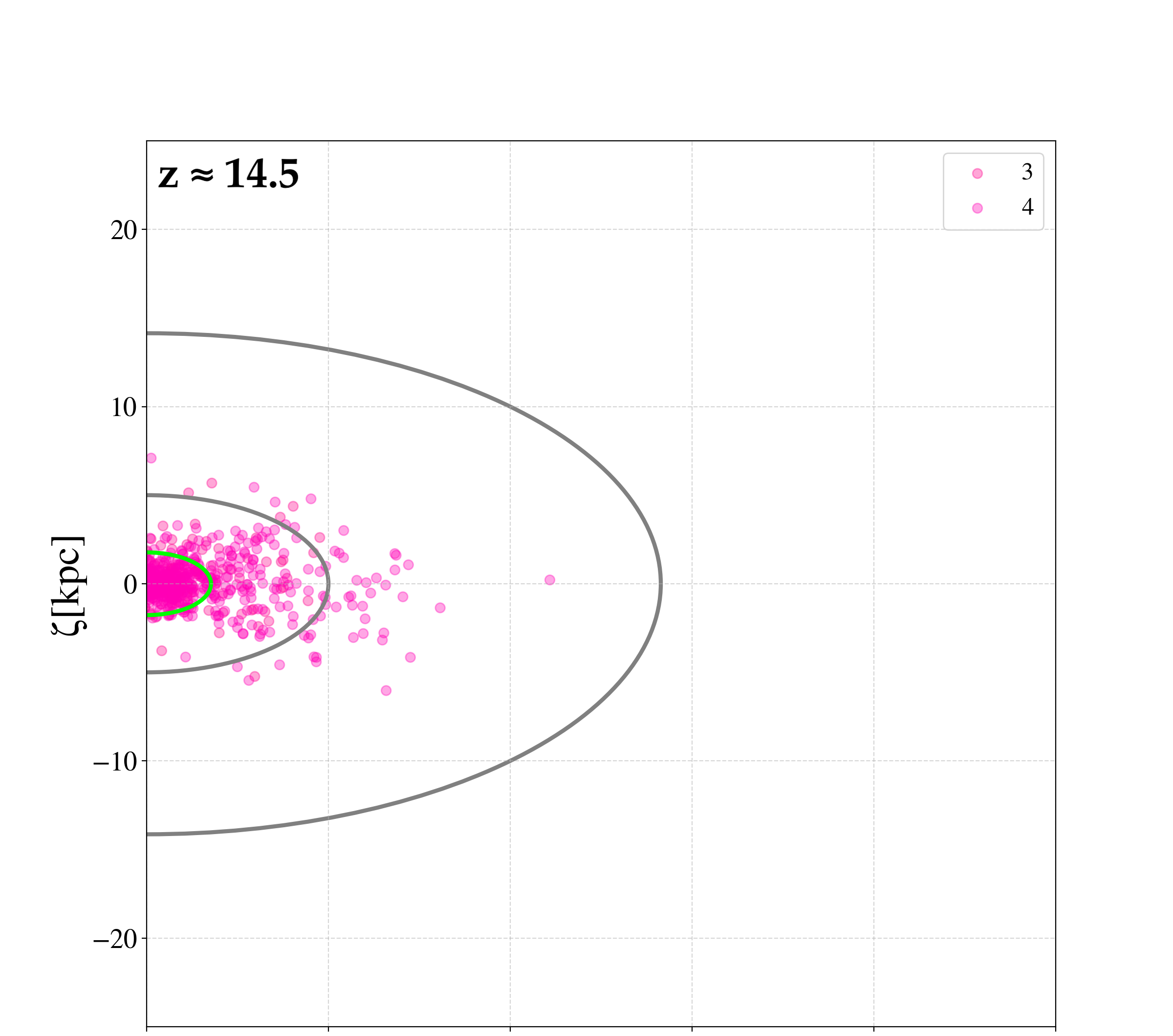} \par \vspace{-22pt}
\includegraphics[width=.75\linewidth]{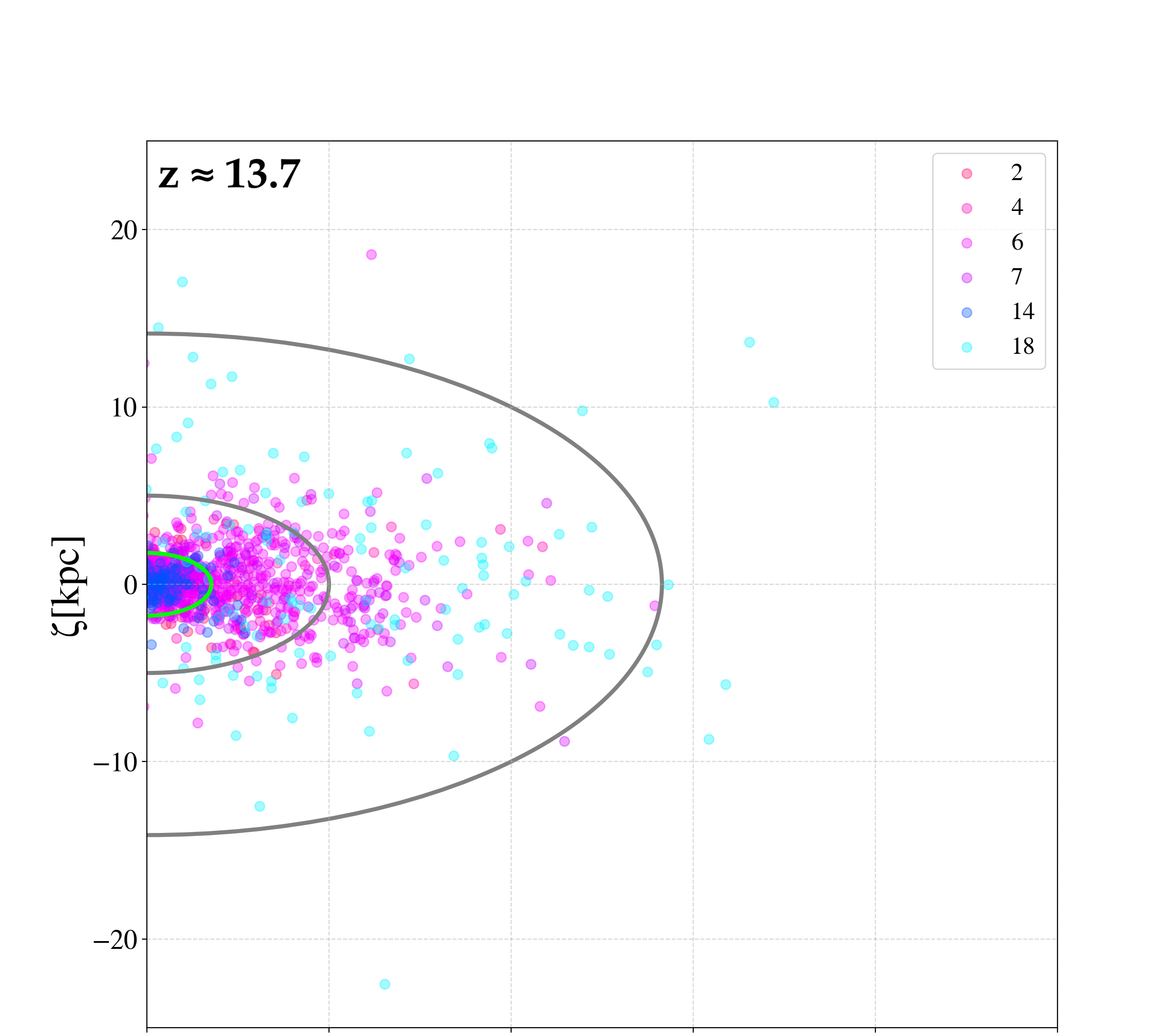}\par \vspace{-22pt}
\includegraphics[width=.75\linewidth]{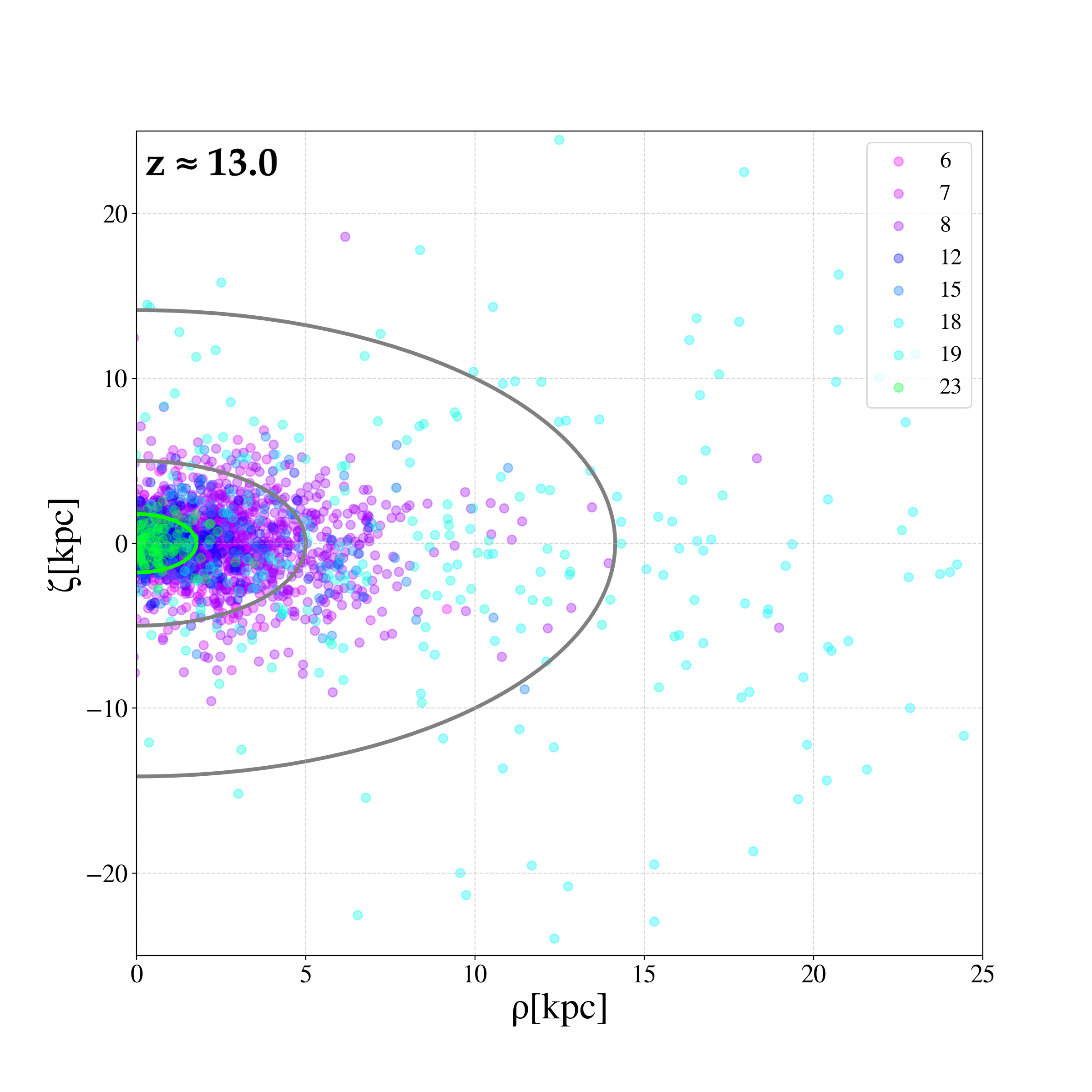} 
\caption{\textit{Present-day} positions, $\rm z = 0$, in the $\rho - \zeta$ cylindrical coordinate plane of Pop~II stars formed in a specific halo of our simulation (id number and color in label), at the specified $z$ which decreases from the top to the bottom panel. Identified Galactic regions are: the inner halo, $7$\,kpc$<r\leq20$\,kpc (grey circles); and the bulge, $r\leq2.5$\,kpc (green circle).}
\label{fig:halos_z}
\end{centering}
\end{figure}

%\begin{figure}
%   \centering
%    \includegraphics[width=0.35\textwidth]{images/halos_maps.jpeg}\hfill%
%    \caption{\textit{Present-day} positions, $\rm z = 0$, in the $\rho - \zeta$ cylindrical coordinate plane of Pop~II stars formed in a specific halo of our simulation (id number and color in label), at the specified $z$ which decreases from the top to the bottom panel. Identified Galactic regions are: the inner halo, $7$\,kpc$<r\leq20$\,kpc (grey circles), and the bulge, $r\leq2.5$\,kpc (green circle).
%    Note that the simulation is not exactly centered in the Galactic centre and therefore the bulge is shifted to $\rho \sim 4$ kpc.}
%    \label{fig:halos_z}
%\end{figure}

\begin{figure*}
    \centering
    \setlength\fboxsep{0pt}
    \setlength\fboxrule{0.25pt}
    \includegraphics[width=\textwidth]{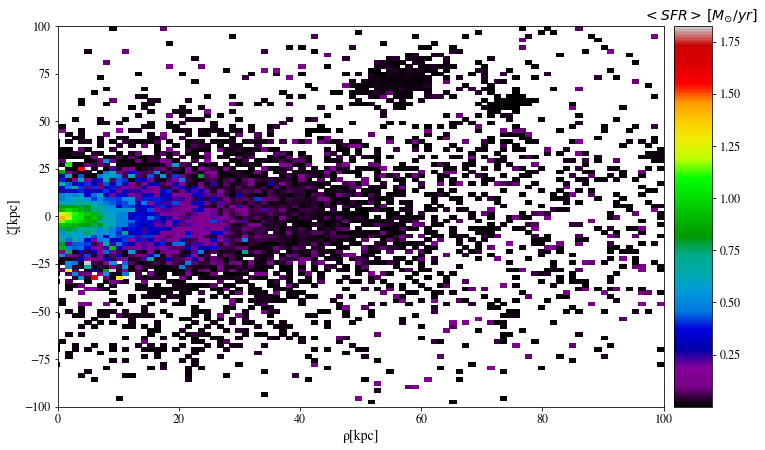}\hfill%
    \caption{Present-day spatial distribution of DM particles in the cylindrical coordinate plane of the simulated galaxy, color-coded with their SFR averaged over the cosmic time between $z=15$ and $z=8$.}
    \label{fig:sfr}
\end{figure*}

\subsection{The first star forming halos}
To identify the first star forming halos and see where the oldest stars are currently located, we reconstructed the hierarchical tree in the MW-analogue by combining the positions of the DM particles within which star formation occurs, and their belonging to a specific halo at different redshifts. A halo is identified as the main progenitor of another at the following timestep, if the latter inherits at least 90\% of its DM particles. If a halo has two or more progenitors then we can assume that a merging process has occurred.

Figure \ref{fig:halos_z} shows the present-day, $z = 0$, positions of Pop II stars formed in the first star-forming halos, at $z \geq 13$. As we can see from the top panel of Fig.~\ref{fig:halos_z}, {\it the oldest Pop\,II stars are currently located in the innermost Galactic region, i.e. the bulge}. The position of DM particles belonging to halos that begin forming stars at lower redshifts (lower panels of Fig.~\ref{fig:halos_z}) is gradually spread towards the outermost regions, filling first the inner and then the outer Galactic halo. These results confirm the idea that the most ancient stellar populations, some of which should have been formed from the ashes of the first stars, must dwell in the bulge (see Sec.~\ref{sec:intro}). 

Figure \ref{fig:sfr} shows the present-day, $z=0$, spatial distribution of DM particles, color-coded with their star-formation rate (SFR) during the first $\approx800$\,Myr, i.e. averaged over the cosmic time between $z=15$ and $z=8$, which corresponds to the lowest redshift for which we have Pop\,III star-formation. Thus we assigned to each DM particle the mean SFR of the halo it belongs to, so that: 
\begin{equation}
\rm \langle SFR\rangle = \frac{\sum_{i=1}^{N}M^{tot}_{*_{i}}}{t(z=8)-t(z=15)},
\label{eq:sfr}
\end{equation}
where $\rm M^{tot}_{*_{i}}$ is the total stellar mass formed in the halo at the {\it i-th} time-step, and $N$ is the number of time-steps between $z=15$ and $z=8$. Then we computed the final $\rm \langle SFR\rangle$ by averaging among the results of all DM particles in the considered pixel. 
This results in a wide range of average star formation rates (Fig.~\ref{fig:sfr}), $\rm \langle SFR \rangle \approx (10^{-4}-1.4)\,M_{\odot}/yr$. 

The spatially resolved region with the highest mean SFR at $\rm z>8$ is the innermost one, i.e.~the Galactic bulge. This result, which is in agreement with previous theoretical \citep[e.g.][]{cescutti2011galactic} and observational \citep[e.g.][]{Lucertini22} findings, suggests that the progenitors of the Galactic bulge experience, on average, a more intense star formation at early times with respect to the progenitors of the Galactic halo.
A higher star formation rate in a metal-poor environment might have strong consequences: in particular, it could allow to produce many rare stellar populations, most likely including the massive progenitors of PISNe \citep{rossi2021ultra}. 

\subsection{Pop\,III enrichment in the first star-forming haloes}

We will now analytically model the Pop\,III star enrichment within the first star forming halos of the cosmological simulation by using more realistic Pop\,III IMFs, and by accounting for their incomplete sampling. To make our analytical calculations, we selected those pristine and star-forming halos for which more than $80\%$ of particles are predicted to dwell in the Galactic bulge, i.e. with $\mathrm{r = \sqrt{\rho^2 + \zeta^2} \leq 2.5\,}\mathrm{kpc}$, at the present time (8~halos). Their halo mass and redshift of formation range between $\rm m_h = (5.5\times10^7\,-\,1.7\times10^8)\,M_{\odot}$, and $\rm 11.4<z<16.3$. 
The predicted star-formation rate for these first star-forming halos is $\rm SFR <1.8\times10^{-2}\,M_{\odot} yr^{-1}$, which in all cases is less than the threshold value for a fully populated Pop\,III IMF, $\rm SFR _{min}\sim 10^{-1} M_{\odot} yr^{-1} $ \citep{rossi2021ultra}. Thus, it is fundamental to include incomplete sampling of the adopted Pop\, III IMFs.

\subsubsection{Realistic IMFs of Pop\,III stars}
\label{sec:imfsampling}

In addition to the delta functions considered so far (Sec.\ref{sec:extreme}), we will assume that Pop\,III stars form according to a {\it Larson IMF}:
\begin{equation}
    \Phi(m) = \frac{dN}{dm} \propto m^{-2.35} \mathrm{exp}(-m_{ch}/m),
    \end{equation}
 biased towards more massive stars as it is expected for the first stellar generations. In particular, following the latest data-driven results from stellar archaeology \citep{rossi2021ultra}, we assume a minimum mass of Pop\,III stars equal to $\rm m_{min}=0.8\,M_{\odot}$, a maximum mass $\rm m_{max}=1000\,M_{\odot}$, and we explore different characteristic masses, $\rm m_{ch} \in \{1, 10, 100 \,M_{\odot}\}$. This allows us to account for the contribution to chemical enrichment of Pop\,III stars exploding as both faint SNe and PISNe and to vary their relative proportions. All the Pop\,III IMFs considered in this study are shown in Fig. \ref{fig:IMF_popIII}.

\begin{figure}
   \centering
    \includegraphics[width=0.48\textwidth]{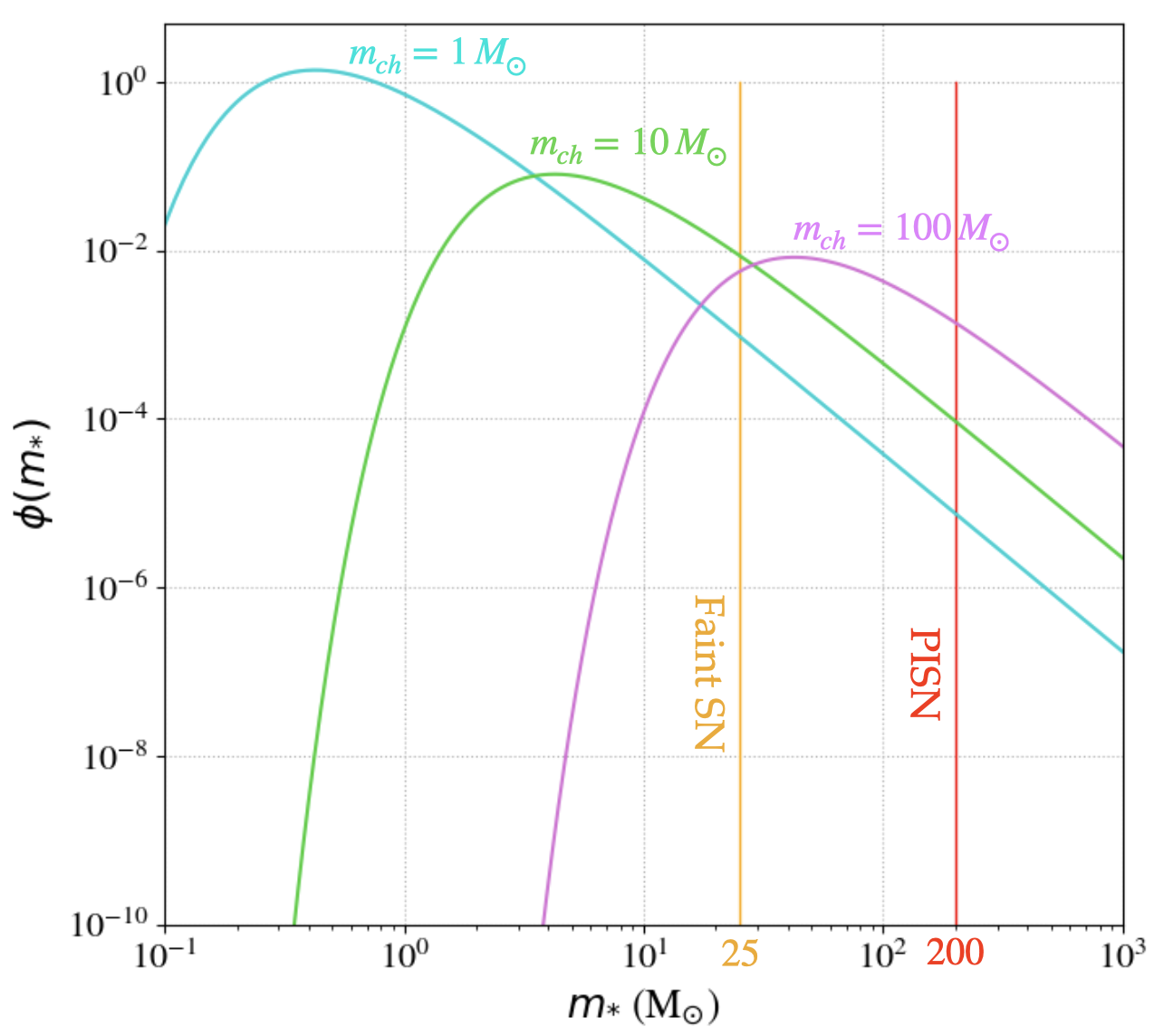}\hfill%
    \caption{The Pop\,III IMFs used in this work: delta functions centered on the average mass values of faint SNe ($\rm m_{popIII}=25\,M_{\odot}$; orange) and PISNe ($\rm m_{popIII}=200\,M_{\odot}$; red), Larson IMFs with $\rm m_{min}=0.8\,M_{\odot}$, $\rm m_{max}=1000\,M_{\odot}$ and $\rm m_{ch} \in \{1, 10, 100 \,M_{\odot}\}$ are respectively in cyan, green and magenta.
    }
    \label{fig:IMF_popIII}
\end{figure}

\begin{figure*}
\centering
\setlength\fboxsep{0pt}
\setlength\fboxrule{0.25pt}
\includegraphics[width=\textwidth]{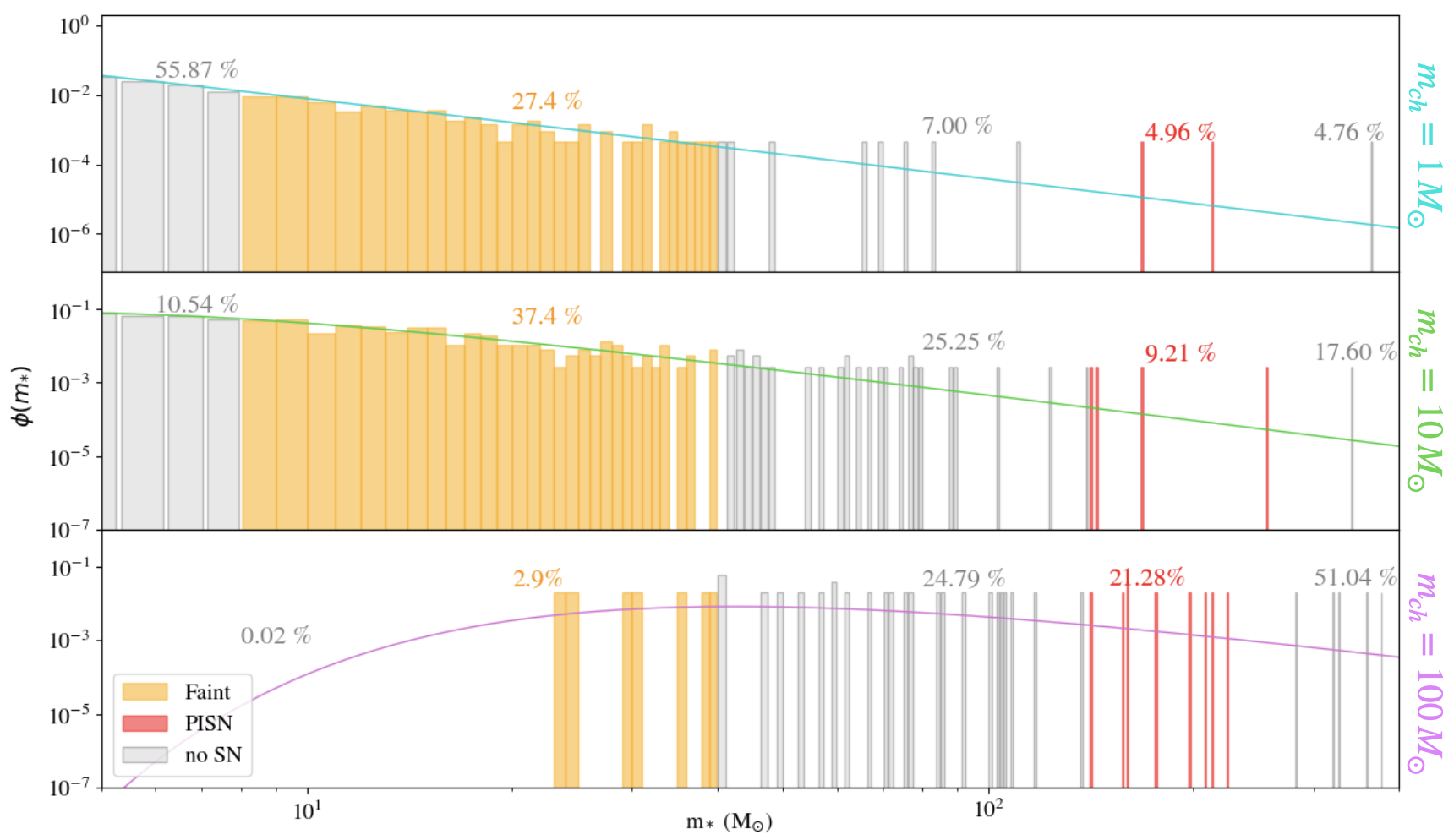}\hfill%
\caption{Theoretical (solid lines) and effective (coloured histogram) Larson IMFs of Pop\,III stars with increasing characteristic masses from top to bottom ($\rm m_{ch} = 1,10,100\,M_{\odot}$) obtained from one run of the random sampling. In each panel, the mass range covered by faint SNe, $\rm (8-40)\,M_{\odot}$ (orange); PISNe, $\rm(140-260)\,M_{\odot}$ (red); and other stars that do not end their lives as supernovae (grey), are specified. The total mass fractions are also listed for each mass range. The total mass of the burst of Pop~III is $7.7\times10^3\,M_{\odot}.$}
\label{fig:imfs_1run}
\end{figure*}

\begin{figure*}
\centering
\setlength\fboxsep{0pt}
\setlength\fboxrule{0.25pt}
\includegraphics[width=0.95\linewidth]{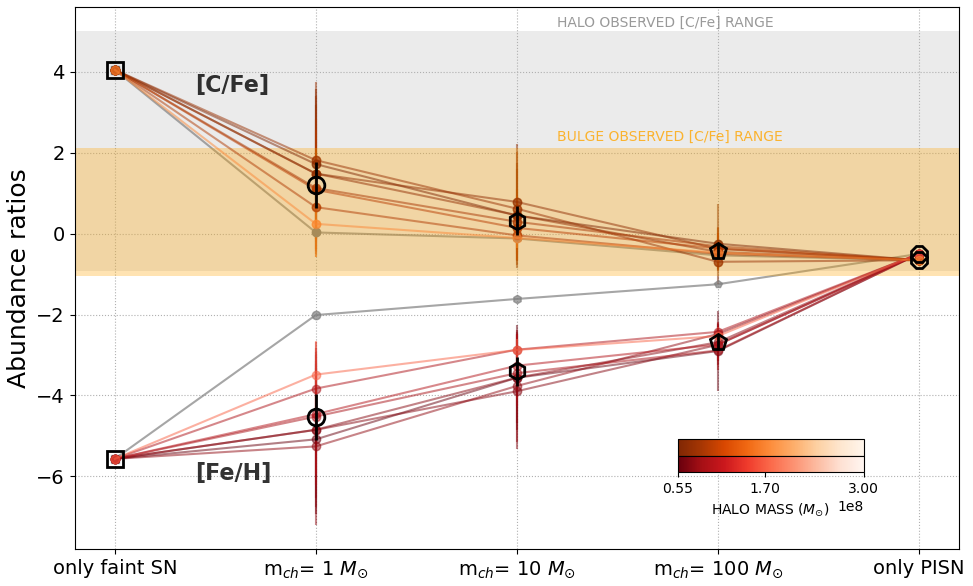}%\hfill%
\caption{Average \feh and \cfe of the ISM in 8 first star-forming halos (colour-coded according to the halo mass), obtained through the random sampling of different Pop\,III IMFs. The characteristic masses ($\rm m_{ch}$) are in solar masses ($M_{\odot}$) and error bars are 1$\sigma$ confidence intervals. The values are computed assuming a time step $\Delta t = 1$ Myr of star formation, consistent with \citet{rossi2021ultra}. For comparison, the grey solid lines are related to the fully sampled IMFs. For each $\rm m_{ch}$ the average abundances over all halos are shown as empty black symbols together
with their errors. Observed ranges of \cfe within the Galactic halo (grey) and bulge (orange) are shown as shaded areas.}
\label{fig:abund_1myr}
\end{figure*}

For each of our selected first star-forming halo, we derived the {\it effective} Pop\,III IMF associated to each burst of primordial star-formation to account for the stochastic and incomplete IMF sampling. This was done using the Monte Carlo procedure developed by \cite{rossi2021ultra}. Starting from the total mass of Pop\,III stars formed, we generate a random number of stars, which are distributed according to the assumed stellar IMF. Due to the stochastic nature of this sampling, every time that stars are formed, the effective stellar mass distribution is differently populated, especially at the higher masses (Fig.~\ref{fig:imfs_1run}). Following \citet{rossi2021ultra} we assumed a time-scale for star formation equal to $\Delta t = 1$ Myr and then we computed the total mass of Pop\,III stars formed in the burst accordingly, $\rm M^{burst}_{popIII}=SFR\times \Delta t$. Then we applied a statistical approach by averaging among the results of $50$ incomplete random sampling of the IMF and quantifying the scatter among them. %In this way we are dividing the mass of the starburst obtained from the simulation (whose timestep is 22 Myr at the considered redshifts) into 22 sub-regions that form stars in 1 Myr.

%For this reason we have applied a statistical approach by averaging among the results of {$\approx 20$} incomplete random sampling of the IMF and quantifying the scatter among them. The reason why we have chosen to average on 22 random sampling is due to our choice of the star formation $\Delta t = 1$ Myr which is consistent with \cite{rossi2021ultra}. In this way we are dividing the mass of the starburst obtained from the simulation (whose timestep is 22 Myr at high redshifts) into 22 sub-regions that form stars in 1 Myr.

Figure \ref{fig:imfs_1run} shows, for different $\rm m_{ch}$, the comparison between the theoretical Pop\,III IMFs and effective ones, which have been obtained in one run of the random sampling. The mass range covered by faint SNe, $\rm (8-40)\,M_{\odot}$; PISNe, $\rm(140-260)\,M_{\odot}$; and other stars that do not end their lives as supernovae are identified using different colours. Among these, stars with $\mathrm{m<8\,M_{\odot}}$ lose their external envelope during the AGB phase, and stars in the two mass ranges, $(\mathrm{40-140)\,M_{\odot}}$ and $(\mathrm{260-1000)\,M_{\odot}}$ are predicted to directly end their life collapsing in a black hole \citep[][]{heger2002nucleosynthetic}. Note that the average mass fraction of stars exploding as faint SNe (and PISNe) strongly varies, not only with $\rm m_{ch}$ but also as a consequence of the incomplete IMF sampling. In the mass range of faint SNe, the Pop\,III IMFs are almost completely sampled except in the extreme case $\rm m_{ch}=100\,M_{\odot}$. On the contrary, in the typical mass range of PISNe the Pop\,III IMF is only partially populated. As expected, however, the mass range of PISNe becomes more densely populated as the characteristic mass increases.

The mass of metals, iron, and carbon produced by the total number of Pop\,III stars formed was computed by summing up the contribution of Pop\,III stars with different masses, i.e. $\rm M_{X} = \sum_{i}^{N} {Y_{X}(m_{popIII,i}){m_{popIII, i}}}$, where $\rm Y_{X}(m_{popIII, i})$ is the yield of the element X produced by Pop\,III stars in the $\rm i-th$ mass bins. For faint SNe, we followed \citet{deBen2016limits} and assumed that the yield corresponding to $25\,M_{\odot}$ is simply re-scaled to the mass of the Pop\,III star exploding as faint SN, i.e. $\rm Y_X(m_{popIII}) = Y_X(25\,M_{\odot}) \times (m_{popIII}/25\,M_{\odot})$. This scaling is quite a good approximation as shown by \citet{marassi2014origin} and it has been used in different works investigating stellar nucleosynthesis \citep[e.g.][]{deBen2016limits, nomoto2013nucleosynthesis}. For PISNe we exploited the yields and SN explosion energies provided by \citet{heger2002nucleosynthetic}. Similarly, we computed the total amount of energy released by SNe by accounting for the number of Pop\,III stars effectively formed in different mass bins, $\rm E_{SN} = \epsilon_w \sum_{i}^{N} {N_{SN}(m^{popIII}_i)\times E_{SN}(m^{popIII}_i)}$, as done in \citet{rossi2021ultra}. %{\bf Stef: Is correct what I wrote?}
Finally, by exploiting Eq.~\ref{eq:m_gas} we computed the ISM metallicity along with \feh and \cfe after the Pop\,III star enrichment. 

\subsubsection{Average and maximum [C/Fe] of bulge stars}

In Fig.~\ref{fig:abund_1myr} we show the \feh and \cfe of the ISM in the 8 first star-forming halos currently dwelling in the Galactic bulge as obtained with our incomplete IMF sampling procedure (Sec.~\ref{sec:imfsampling}) and as a function of the five different Pop\,III IMFs assumed (see Fig.~\ref{fig:IMF_popIII}). The abundance ratios for the different halos are colour-coded according to the halo mass and, for comparison, we also show the results obtained considering the fully sampled IMFs as solid grey lines. 
For each $\rm m_{ch}$ the average abundances over all halos are shown as empty black symbols together with their errors.
The observed ranges of \cfe within the Galactic halo and bulge (from Fig.~\ref{fig:fcemp_halo}) are highlighted as well in Fig.~\ref{fig:abund_1myr}  (respectively as grey and orange shaded areas).

First, we note that for all IMFs, the total gas metallicity after the Pop\,III star enrichment is $\rm Z_{ISM} > Z_{cr} = 10^{-4.5}Z_{\odot}$, which implies that normal low-mass long-lived stars will be able to form in such environments. Furthermore, when we only account for the contribution of faint SNe, we get a very low value of \feh $\simeq -5.6$ and an extremely high $\cfe\simeq+4$, which is close to the maximum value observed in the Milky Way halo \citep{Keller2014}. As soon as the chemical contribution of PISNe is also considered, namely for $\rm m_{ch}=1\,M_{\odot}$, the \cfe value drops dramatically by at least 2 orders of magnitude. As we can see at fixed $\rm m_{ch}$, as the halo mass increases, the abundances approach the fully sampled case, that is, \feh increases and \cfe decreases. This result can easily be explained as a consequence of the Pop\,III IMF sampling: in the lower mass halos, which have lower star formation rates, on average only one PISN explodes, thus only partially lowering the \cfe value obtained in the case of faint SNe only. Still, even a single PISN is able to inject into the ISM $50\%$ of its total stellar mass in form of heavy elements, thus strongly affecting the final [C/Fe] value of the ISM. In the most massive halo more PISNe can actually form and explode, thus further lowering the expected \cfe value. For increasing characteristic mass, the mean \cfe values become even lower, since PISN production starts to dominate the chemical enrichment. When only PISNe explode, no CEMP stars are able to form, regardless of the halo mass.

In conclusion, the measured \cfe values in the Galactic bulge suggest that massive Pop\,III stars exploding as PISNe have likely formed in this environment, washing out the high \cfe signature left by low-energy primordial SNe. Furthermore, by comparing our predicted \cfe with the values observed in the bulge (Fig.~\ref{fig:abund_1myr}), we can exclude the two extreme IMF cases, where Pop\,III stars explode either only as faint SNe or only as PISNe, and suggest that the IMF that best reproduces the observations is a Larson type with a characteristic mass possibly $\rm 1\,M_{\odot} \lesssim m_{ch}\lesssim 10 \,M_{\odot}$. This result is in full agreement with cosmological simulations for the formation of Pop~III stars \citep[e.g.][]{hirano2014one,Susa2014,hirano2015} or theoretical models interpreting different observables \citep[e.g.][]{deBen2016limits,rossi2021ultra,Ishigaki2018}. %$\rm m_{ch}\gtrsim1 \,M_{\odot}$.

\section{Summary and Conclusions}
\label{sec:concl}
The aim of this paper is to investigate the apparent dearth of carbon-enhanced metal-poor (CEMP) stars with high \cfe values in the Galactic bulge with respect to other environments, such as the Galactic halo and ultra-faint dwarf galaxies \citep[e.g.][]{Howes2015,howes2016embla,arentsen2021pristine}. This lack is particularly puzzling since the bulge is supposed to be the oldest Galactic stellar component \citep{white2000first,diemand2005earth,tumlinson2009chemical,salvadori2010mining,starkenburg2016oldest} and CEMP stars are predicted to be among the most ancient observable stars, most likely being the direct descendants of first stars with intermediate masses that exploded as low-energy faint supernovae \citep[e.g.][]{iwamoto2005first,marassi2014origin,bonifacio2015topos,deBen2016limits}. A reason for the dearth of CEMP stars in the Galactic bulge could be linked to a bias introduced by the photometric selection performed by the surveys that have targeted metal-poor stars in this environment \citep[see Sec.~\ref{sec:intro} and][for details]{Howes2015,howes2016embla,arentsen2021pristine}. Despite these caveats, this issue still persists even though progress has been achieved over the years in detecting these types of stars leading to an actual increase in the number of their observations. Therefore, we asked ourselves:
could this scarcity of CEMP stars be a consequence of the low statistics of metal-poor stars in the more metal-rich and dusty bulge? Could it suggest an intrinsically different formation mechanism of this region compared to the other environments?

In order to answer these questions, we carried out a diversified investigation which is summarised in the following points:

\begin{itemize}

\item We first performed a statistical analysis of the metal-poor stars observed in the Local Group to understand whether the dearth of CEMP stars is due to the limited sample size of metal-poor stars within the bulge, which is dominated by metal-rich stars \citep[e.g.][]{zoccali2008metal,ness2013argos,howes2016embla}. By exploiting available data (Sec.~\ref{sec:obs}), we first derived the fraction of CEMP stars at a given \feh for the Galactic halo ($F^{halo}_{\mathrm{CEMP}}$), which is the environment with the highest statistic on very metal-poor stars, $\feh<-2$. We then assumed this fraction to be {\it ``Universal''} and combined it with the MDFs obtained for different dwarf galaxies and for the bulge. As a final step we computed the probability of identifying a CEMP star in a given \feh range, $P_{\mathrm{obs}}$, while blindly observing these environments. In agreement with previous studies \citep{salvadori15}, we show that $P_{\mathrm{obs}}$ is strongly affected by the total stellar mass and average metallicity of the examined environment. In the bulge, which is the brightest and most metal-rich, we find that $P_{\mathrm{obs}}<0.01\%$, i.e. it is more than two orders of magnitude lower with respect to ultra-faint dwarf galaxies.\\ 

We conclude that CEMP stars could be partially hidden in this region dominated by metal-rich stars, rather than being totally absent. However, this dearth cannot totally be ascribed to this statistical reason. If the CEMP fraction in the bulge would be the same as in the halo, then biased surveys specifically searching for the most metal-poor stars in the Galactic bulge - such as the PIGS survey \citep[][]{arentsen2021pristine} - should have observed many more CEMP stars ($\sim450$) than what they actually found ($<62$, Sec.~\ref{sec:obs}). This discrepancy could be due to selection biases in the photometric selection, but could there be another mechanism that reduces the \cfe values in the bulge?  %Furthermore, we cannot explain why there should be a preferential dearth of CEMP stars with high [C/Fe] values.
\\

\item We then asked whether the lower \cfe values of CEMP stars in the Galactic bulge reflects the different formation and evolution of this ancient environment. To address this question, we focused on the predictions derived from the $\Lambda$CDM cosmological model, through the use of a $N$-body simulation that follows the hierarchical formation of a MW-like galaxy combined with the semi-analytical model {\tt GAMETE} \citep[e.g.][]{salvadori2007cosmic,salvadori2010mining,salvadori15,graziani2015galaxy,pacucci2017gravitational}, which is required to follow the star formation and metal enrichment history. The model is data-calibrated, i.e. the best values of the free parameters are set to reproduce the observed properties of the MW at $z=0$ (Sec.~\ref{sec:model}). If we assume that {\it all} Pop\,III stars explode as faint SNe, we find that the mass fraction of CEMP stars with \cfe$>+2$ increases at decreasing Galacto-centric radii and it is maximum in the Galactic bulge, which is the region of the simulation containing the most ancient stars, which is at odd with observations.
However, the \emph{N}-body simulations reveal that the stars dwelling into the present-day bulge form in halos that experienced the {\it highest mean star-formation rate} at high-redshifts ($z > 8$).\\ 

We inferred that the dearth of CEMP stars with \cfe$>+2$ in the Galactic bulge might be linked with the higher star-formation rate of its early progenitor halos, which hosted the first stars. Indeed, star-formation rates  $>10^{-2}M_{\odot}/yr$ in primordial environments might allow the formation of rare very massive Pop\,III stars \citep{rossi2021ultra}, which evolve as energetic Pair Instability SNe (PISNe).\\ 

\item We thus investigated how the chemical enrichment of the bulge progenitors depends upon the properties of rare Pop\,III stars that can {\it effectively form} in these highly star-forming systems. To this aim, we performed analytical calculations and computed the chemical properties of the ISM in the bulge progenitors after the contribution of Pop\,III stars. More specifically, we investigated how different Pop\,III IMFs affect the \cfe and \feh values of the ISM by assuming various characteristic masses ($\rm m_{ch}= 1, 10, 100 \,M_{\odot}$), by including the contribution of Pop\,III stars exploding as PISNe (see Sec.~\ref{sec:results}), and by accounting for the incomplete sampling of the Pop\,III IMF \citep{rossi2021ultra}. Our results show that very massive Pop~III stars can effectively form in the bulge progenitors, and that their contribution to the chemical enrichment as energetic PISNe can partially wash out the distinctive signature of faint SNe, lowering the carbon overabundance down to $\cfe < +2$. In particular, we show that the higher the probability to form very massive Pop\,III stars, i.e. the larger the $\rm m_{ch}$, the lower is the $\cfe$ value of the ISM after the contribution of Pop\,III stars. By exploiting the available data we thus tentatively infer $\rm 1\,M_{\odot} \lesssim m_{ch}\lesssim 10 \,M_{\odot}$,
%$\rm m_{ch}\gtrsim 1 \,M_{\odot}$, 
consistent with the constraints found by \citet{rossi2021ultra} based on ultra-faint dwarfs.\\ 

We conclude that the modest [C/Fe] values of CEMP stars identified in the bulge, \cfe$\approx +0.8$, along with the dearth of CEMP stars with \cfe$>+2$ could be an indirect probe of very massive first stars exploding as PISNe, which are extremely rare and hence can only form in highly star-forming progenitors of the MW bulge. 
\end{itemize}

Through careful analysis of observational data and theoretical simulations, we thus suggest that the dearth of CEMP stars in the bulge might be intrinsic, and not only a consequence of systematic observational effects. Furthermore we have shown that the first star forming halos that end up in the bulge have typically significantly higher star-formation rates than those in the outskirts of the galaxy. This means that the Pop~III IMF is better sampled in these systems, enabling the formation of rare populations such as very massive first stars that explode as PISNe. Ultimately, our analysis showed this to be a very plausible explanation for the dearth of CEMP stars in the bulge, as added PISN contribution will lower the observed [C/Fe]. %\stef{If we have a figure for \cfe vs PISN fraction we should discuss it here eventually eliminating the last sentence}. 
Furthermore we suggested a new promising method that exploits the lack of CEMP stars to constrain the characteristic mass of the first stars. %, which with the currently available data favours a top-heavy Pop\,III IMF with $\rm m_{ch}\approx 10 \,M_{\odot}$. 

\section{Discussion and future outlook}
Our findings provide a plausible explanation for the lack of CEMP stars with high \cfe values in the Galactic bulge, which is not only linked to statistical issues or observational biases, but it is rather a consequence of the different formation path of this ancient but star-rich environment. The idea that Pop\,III stars exploding as PISNe can efficiently form in the bulge progenitors and lower the [C/Fe] value in their ISM is extremely appealing, since it can also explain the larger fractions of CEMP stars observed in the Galactic halo and in UFDs. The typical star-formation rates observationally inferred for UFDs ($\leq 10^{-3}M_{\odot}$/yr, e.g., \citealt{salvadori2014, Gallart2021}) are indeed too low for enabling these galaxies to form very massive Pop\,III stars \citep{rossi2021ultra}. 
%STEF: TO BE ADDED IN THE BIBTEXx
%https://ui.adsabs.harvard.edu/abs/2014MNRAS.437L..26S/abstract
%https://ui.adsabs.harvard.edu/abs/2021ApJ...909..192G/abstract
The early chemical enrichment stages of these small systems are thus likely dominated by faint SNe, whose explosion energy ($E < 10^{51}$~erg) is small enough to not exceed the halo binding energy (Rossi et al. in prep) and whose chemical products are characterized by a large amount of carbon and a tiny amount of iron. The progenitors of UFDs have been suggested to build-up the low-Fe tail of the Galactic halo \citep[e.g.][]{salvadori15,bonifacio2021topos}, where CEMP-no stars are found, making the overall scenario perfectly consistent. 

Other sources that can partially wash-out the chemical signature of faint SN are primordial hypernovae, i.e. Pop\,III stars with masses ${\rm m_{PopIII}\sim (10-100) M_{\odot}}$ that experience energetic SN explosions, $E \sim 10^{52}$~erg \citep{HegerWoosley2010}. In the last years the descendants of primordial hypernovae have been firstly identified thanks to their sub-solar [C/Fe] values in nearby dwarf spheroidal galaxies and halo stars \citep{Skuladottir21,Placco21}. However, since Pop\,III stars exploding as faint SNe and as hypernovae have the same range of masses, we expect them to form with the same efficiency in both the progenitors of the Galactic halo, UFDs, and the bulge. Furthermore, the same reasoning can be applied to the contribution of Pop~II stars in the chemical enrichment. In conclusion, this alternative solutions to explain the lower \cfe value of bulge stars, is hard to reconcile with the global scenario. In a forthcoming study, we will further explore the contribution of different Pop~III stars to the bulge enrichment by performing a self-consistent calculation, i.e. by following within the $N$-body simulation the chemical elements produced by Pop~III stars with different masses and explosion energies (Koutsouridou et al. in prep.). In the same study we will relax the Instantaneous Recycling Approximation, which, although quite robust for $\rm z>6$, may have affected our results. In fact, taking into account the lifetimes of Pop III stars, the more massive PISNe would have exploded first - raising Fe and keeping C low - and then the faint SNe would have contributed - raising C with constant Fe. Consequently among the older Pop II stars we could have stars with higher \feh and lower \cfe.

Ultimately, the dearth of CEMP stars with high \cfe values in the Galactic bulge might provide an {\it indirect} probe for the long-searched very massive Pop\,III stars evolving as PISNe. A way to prove that primordial PISNe exist is to search for an under-abundance of [Zn/Fe] and [Cu/Fe] in their descendants \citep{salvadori2019probing, aguado23b}. Interestingly, stars with sub-solar [Zn/Fe] have been predominantly identified in the bulge \citep[e.g.][]{barbuy2015zinc, Duffau17} and classical dwarf spheroidal galaxies \citep[e.g.][]{Skuladottir17}, which are indeed more massive and star-forming than UFDs.
%https://ui.adsabs.harvard.edu/abs/2017A%26A...606A..71S/abstract 
Furthermore, in the nearby future the WEAVE \citep{dalton2016weave}, 4MOST \citep{Christlieb19, chiappini20194most, bensby20194most}, and J-PLUS/S-PLUS \citep[][]{cenarro2019j,mendes2019southern} Galactic Archaelogy surveys will allow us to derive a large number of carbon measurements in both halo and bulge metal-poor stars, as well as the satellite dwarf galaxies (4DWARFS; Skúladóttir et al. 2023, in press.). Thus, the CEMP-no frequency will be established with unprecedented accuracy, giving new insights into PISN events. Unfortunately, Zn will not be measured by 4MOST. However, if the green grating will be used in the high-resolution WEAVE survey, a large sample of Cu and Zn measurements will be extremely valuable to confirm this scenario and to search for PISNe descendants.

This is the perfect time to deepen our studies on Stellar Archaeology. The future of the field is indeed promising since great technology developments and new instruments will provide us unprecedented datasets to confirm our theoretical findings. In the coming years, there will be a surge of observations in the
inner Galaxy of ancient, metal-poor stars 
thanks to the 4MOST Milky Way Disc and BuLgE Low-Resolution \citep[4MIDABLE-LR,][]{chiappini20194most} and 4MOST MIlky Way Disc And BuLgE High-Resolution surveys \citep[4MIDABLE-HR,][]{bensby20194most}. This will provide a completely new view of the central regions of the Milky Way, and, hopefully, greatly advance our understanding of the first stars.

%If our PISNe scenario for the dearth of CEMP stars within the bulge will be confirmed, then we will be able to constrain the characteristic mass of the Pop\,III IMF by exploiting the average/maximum observed \cfe ranges of bulge stars.  
%{\bf Stef}: Add a last sentence on the Pop~III IMF ?

\section*{Acknowledgements}
SS, MR, DA, IK, and \'{A}S acknowledge funding from the European Research Council (ERC) under the European Unionion's Horizon 2020 research and innovation programme, project NEFERTITI (grant agreement No 804240). SS also acknowledges funding from the PRIN-MIUR17, The quest for the first stars, prot. n. 2017T4ARJ5. The authors acknowledge the SDSS/APOGEE survey.

%%%%%%%%%%%%%%%%%%%%%%%%%%%%%%%%%%%%%%%%%%%%%%%%%%
\section*{Data Availability}

The authors confirm that the data analysed in this study are available within the JINABASE \citep[][]{abohalima2018jinabase}, the APOGEE spectroscopic survey \citep[DR16,][]{jonsson2020apogee}, 
and the following papers: \citet{Norris_2010}, \citet{Frebel_2014}, \citet{refId0}, \citet{Frebel_2009}, \citet{Simon_2010}, \citet{Lai_2011}, \citet{Norris_2010}, \citet{Gilmore_2013}, \citet{Kirby_2015}, \citet{Frebel2010}, \citet{tafel_2010}, \citet{Starkenburg_2014}, \citet{Simon_2015}, \citet{Jablonka_2015}, \citet{skuladottir2015}.
PIGS data have been provided via private communication with Anke Arentsen.
The theoretical data that support the findings of this study are\ available upon request to the corresponding author GP.



\bibliographystyle{mnras}
\bibliography{example} % if your bibtex file is called example.bib

%%%%%%%%%%%%%%%%%%%%%%%%%%%%%%%%%%%%%%%%%%%%%%%%%%

%%%%%%%%%%%%%%%%% APPENDICES %%%%%%%%%%%%%%%%%%%%%

%\appendix



%%%%%%%%%%%%%%%%%%%%%%%%%%%%%%%%%%%%%%%%%%%%%%%%%%

% Don't change these lines
\bsp	% typesetting comment
\label{lastpage}
\end{document}